\documentclass[a4paper,11pt]{article}

\pdfoutput=1

\usepackage{jcappub}
\usepackage{amsmath}
\usepackage{amsfonts}
\usepackage{amssymb}
\usepackage{mathtools}
\usepackage{graphics}
\usepackage{citesort}
\usepackage{graphicx}
 \usepackage{url}

\usepackage[utf8]{inputenc}

\usepackage{hyperref}

\title{Lifting the core-collapse supernova bounds on keV-mass sterile neutrinos}

\author[a]{Anna M.~Suliga,}
\emailAdd{anna.suliga@nbi.ku.dk}
\author[a]{Irene Tamborra,}
\emailAdd{tamborra@nbi.ku.dk}
\author[b,c,d]{Meng-Ru Wu}
\emailAdd{mwu@gate.sinica.edu.tw}

\affiliation[a]{Niels Bohr International Academy and DARK, Niels Bohr Institute, University of Copenhagen, Blegdamsvej 17, 2100, Copenhagen, Denmark}
\affiliation[b]{Institute of Physics, Academia Sinica, Taipei, 11529, Taiwan}
\affiliation[c]{Institute of Astronomy and Astrophysics, Academia Sinica, Taipei, 10617, Taiwan}
\affiliation[d]{Physics Division, National Center for Theoretical Sciences, Hsinchu, 30013, Taiwan}

\abstract{We explore the energy and entropy transport as well as the   lepton number variation induced from the mixing between electron and sterile neutrinos with keV mass in the supernova core.  We develop a radial- and time-dependent treatment of the $\nu_s$--$\nu_e$ mixing, by including 
ordinary matter effects, reconversions between sterile and electron antineutrinos, as well as the collisional production of sterile particles for the first time. The dynamical feedback due to the production of sterile particles on the composition and thermodynamic properties of the core  only leads to major implications for the supernova physics for large mixing angles ($\sin^2 2\theta \gtrsim 10^{-10}$). Our findings suggest that a self-consistent appraisal of the electron-sterile conversion physics in the supernova core would relax the bounds on the sterile neutrino mixing parameters reported in the literature for $\sin^2 2\theta \lesssim 10^{-6}$, leaving  the $(m_s, \sin^2 2\theta)$ parameter space relevant to dark matter searches unconstrained by supernovae. 
}

\definecolor{green}{rgb}{0,0.5,0}
\begin{document}

\maketitle

\section{Introduction}
\label{sec:Introduction}
The physics of weak interactions, and in particular of neutrinos,  is known to be of tremendous importance to the core-collapse supernova (SN) mechanism~\cite{Mirizzi:2015eza,Janka:2012wk,Janka:2017vcp}. However, the behavior of neutrinos in core-collapse SNe is still surrounded by large uncertainties, given the major conceptual and numerical challenges that it poses~\cite{Mirizzi:2015eza,Chakraborty:2016yeg}. 
This already involved scenario may be further complicated if particles beyond the Standard Model exist.
An example, in this context, may be coming from sterile neutrinos, whose existence is compatible with terrestrial data, as well as  cosmological and astrophysical surveys, and may explain the origin of neutrino masses  and be good dark matter candidates~\cite{Merle:2017dhf,Boyarsky:2018tvu,Abazajian:2012ys,Abazajian:2019ejt}.

If heavy sterile neutrinos with $\sim$~keV--MeV mass  exist, preliminary work shows that they could have a dramatic impact on the SN engine~\cite{Shi:1993ee,Nunokawa:1997ct,Abazajian:2001nj,Hidaka:2006sg,Hidaka:2007se,Raffelt:2011nc,Arguelles:2016uwb,Suliga:2019bsq,Warren:2014qza,Warren:2016slz,Mastrototaro:2019vug,Syvolap:2019dat,Rembiasz:2018lok}.
Given the major challenges already existing for  the modeling of standard neutrino  physics in the SN core, the sterile neutrino problem in SNe is far from being solved. 
 In Ref.~\cite{Suliga:2019bsq}, we recently focused on sterile neutrinos with mass between 1 and 100~keV and, for the first time, attempted a radial and time dependent modeling of the sterile neutrino mixing with $\tau$ neutrinos by taking into account the collisional production of sterile particles as well as their matter-enhanced production. By consistently implementing the feedback on the  
 neutrino lepton number, we found that a large $\nu_\tau$--$\bar\nu_\tau$ lepton asymmetry can develop. Our findings highlighted the importance of the dynamical feedback of the flavor conversion physics on the SN background and hinted towards a region of the mass-mixing parameter space, possibly excluded by SNe, smaller than previously estimated.

 The $\nu_e$--$\nu_s$ mixing has been investigated within a simplified framework in the past~\cite{Hidaka:2006sg,Hidaka:2007se,Warren:2014qza,Warren:2016slz,Shi:1993ee,Nunokawa:1997ct,Abazajian:2001nj}.
 Following up on the pioneering work of Refs.~\cite{Shi:1993ee,Nunokawa:1997ct}, Ref.~\cite{Hidaka:2006sg} explored the $e$--$s$ mixing through a one-zone model by focusing on the  in-fall phase of the SN collapse. A significant reduction in the core lepton fraction was found. The latter may likely result in a smaller homologous core, and therefore  in a smaller initial shock energy, preventing a successful shock re-heating.
 On the other hand,  the $e$--$s$ mixing in the early post-bounce SN phase may be responsible for  transport-enhanced entropy deposition ahead of the shock~\cite{Hidaka:2007se}. In the latter case, the neutrino luminosity enhancement and pre-heating could increase the likelihood of a successful core-collapse  explosion.
 Moving beyond the one-zone modeling, but applying a simplified treatment of the flavor conversion physics, Refs.~\cite{Warren:2014qza,Warren:2016slz} attempted to simulate the sterile neutrino production in a  hydrodynamical simulation of the core collapse. An increase of the kinetic energy due to neutrino reheating behind the shock was found   after $\simeq 200$~ms post-bounce. In addition, the production of sterile particles alters the emission of all three neutrino flavors, leading to increased neutrino luminosities at early times. As a consequence, an increase of the shock energy was also found for a broad range of sterile mixing parameters~\cite{Warren:2014qza}.

Given the potential  importance that the $e$--$s$ mixing could have in SNe, in this paper, we  extend the work of Ref.~\cite{Suliga:2019bsq}. 
For the first time, our work  tracks the flavor evolution by including ordinary matter effects, collisional production of sterile neutrinos, as well as reconversion effects. We  rely on a static hydrodynamical background and, by implementing a radial and time-dependent modeling of the flavor evolution, investigate the dynamical feedback that the production of sterile particles induces on the energy transport, the medium temperature, the entropy, and the chemical potentials of leptons and baryons inside the SN core. Our framework  allows to glimpse on the related impact on the SN dynamics in a more consistent manner than in  previous work.

This paper is organized as follows. Section~\ref{sec:ref_signal} introduces the reference SN model used in the calculations. Section~\ref{sec:conversions} focuses on the physics of neutrino flavor conversions in the SN core. The effect of the dynamical feedback due to the production of sterile particles on the evolution of the particle chemical potentials, as well as  entropy and medium temperature is discussed in Sec.~\ref{sec:Feedback}.  
Section~\ref{sec:SNbounds} focuses on how the SN bounds on the sterile neutrino mixing parameters are affected by our findings. 
The implications on SN explosions enhanced by sterile neutrinos are further
discussed in Sec.~\ref{sec:SNexplosion}. Finally, we summarize our work in Sec.~\ref{sec:Conclusions}. 
Additionally, the computation of the charged-current interaction rates of neutrinos and their related Pauli blocking factors is presented in Appendix~\ref{appendix:Pauli}. Appendix~\ref{sec:app_qcdot} discusses the neutrino heating and cooling in the presence of active-sterile mixing, while Appendix~\ref{appendix:Temperature} illustrates the procedure adopted to compute the medium temperature in the presence of the dynamical feedback.

\section{Benchmark supernova model}
\label{sec:ref_signal}

In order to simulate the sterile neutrino production within a  realistic setup, we rely on a one-dimensional spherically symmetric simulation of a SN with mass of $18.6\ M_{\odot}$~\cite{Bollig2016} and SFHo nuclear equation of state (EoS). As we aim to investigate the impact of the electron-sterile neutrino conversions during the SN accretion phase and the potential impact on the explosion mechanism,  we select the inputs from  one representative post-bounce time snapshot: $t_\mathrm{pb} = 0.25~\mathrm{s}$. 

\begin{figure}[b]
\centering
\includegraphics[scale=0.3]{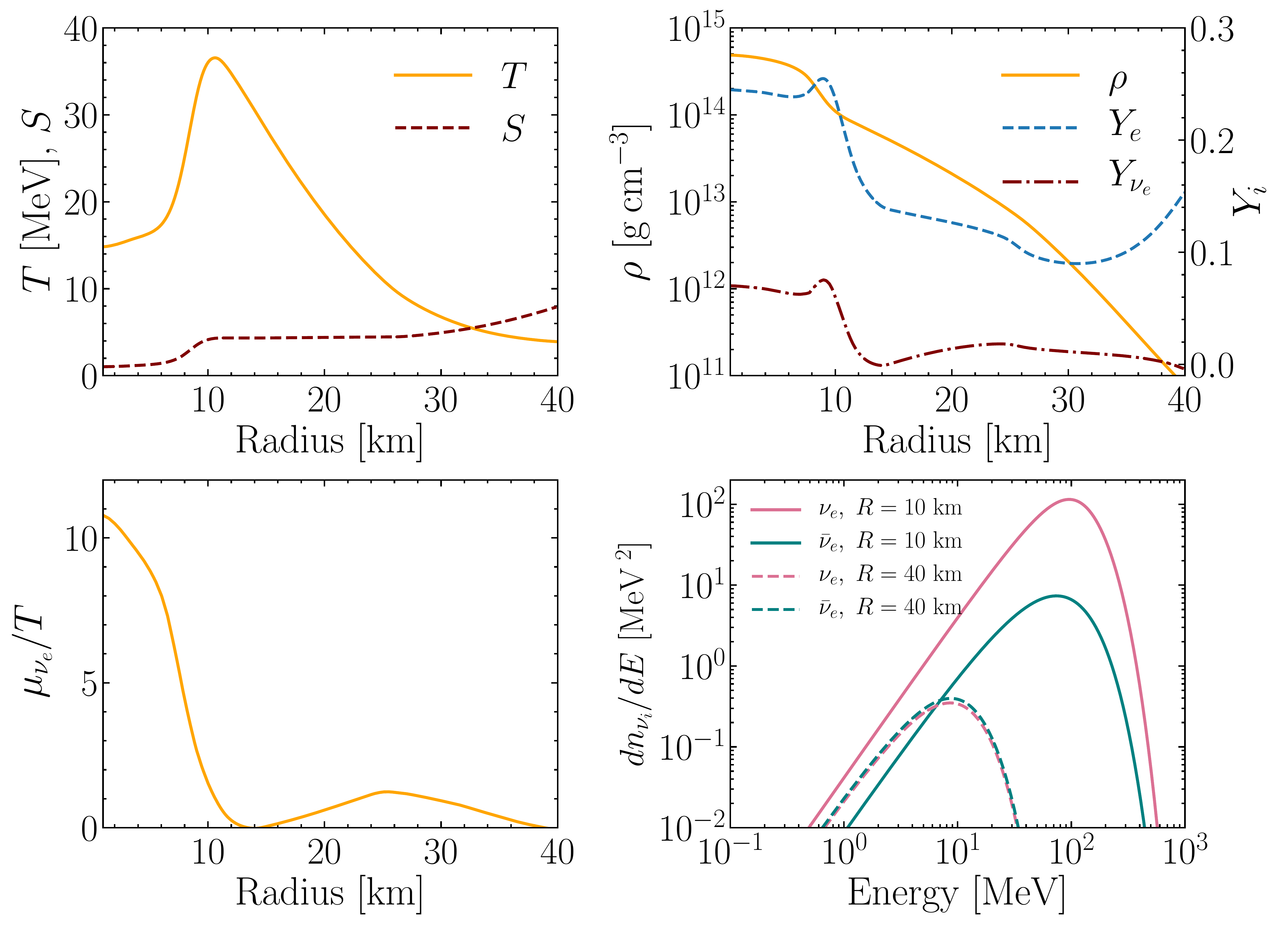}
    \caption{Initial setup employed in the modeling of the sterile neutrino production for our SN benchmark model. {\it Top left:} Radial profile  of the temperature (orange, solid line) and entropy (brown, dashed line). {\it Top right:} Baryon density (orange, solid line), electron fraction (blue, dashed line), and electron neutrino fraction (brown, dash-doted line) as a function of the  star radius. {\it Bottom~left:}~Radial profile of the neutrino degeneracy parameter. {\it Bottom right:} Energy distributions of $\nu_e$ (pink) and $\bar\nu_e$ (teal) as  functions of the neutrino energy at $10$ and $40$~km (solid and dashed lines, respectively) from the SN core.
    }
    \label{fig:Fig1}
\end{figure}
Figure~\ref{fig:Fig1} shows the radial profile of the initial medium temperature and entropy per baryon on the top left panel;  the baryon density, the initial electron and electron-neutrino fractions ($Y_e$ and $Y_{\nu_e}$, respectively) are shown as function of the radius  on the top right panel, and  the bottom left panel of Fig.~\ref{fig:Fig1} illustrates the neutrino degeneracy parameter (e.g.~the ratio between the initial electron neutrino chemical potential $\mu_{\nu_e}$ and the initial medium temperature $T$). 
Note that the maximum radius plotted in Fig.~\ref{fig:Fig1} is the neutrinosphere radius, $R_\nu\simeq 40~\mathrm{km}$. We focus on this spatial region, corresponding to the proto-neutron star, because this is where the active-sterile flavor conversions mainly occur  for the  mass range of the sterile particles explored in this paper (see Sec.~\ref{sec:conversions});  however,  as  will be discussed in Sec.~\ref{sec:MSW}, neutrino flavor conversions may take place only when $\mu_{\nu_e}/T \lesssim 1$, i.e.~the neutrino degeneracy is negligible.
 As we will discuss in Sec.~\ref{sec:Feedback}, all these quantities will be affected by the sterile-electron neutrino conversions in our simulations, except for the baryon density.

In the SN core, $\nu_e$ and $\bar\nu_e$ are trapped and are in thermal equilibrium with the medium. Their initial energy distributions are plotted in 
Fig.~\ref{fig:Fig1} (bottom right panel) for two different distances from the SN core and are well fitted by a Fermi-Dirac distribution:
\begin{equation}
\label{eq:F-D}
\frac{dn_{\nu}}{dE} = \frac{1}{2\pi^2}\frac{E^2}{e^{(E-\mu_\nu)/T} + 1} \ , 
\end{equation}
where $\mu_\nu$ is the flavor-dependent neutrino chemical potential, such that $\mu_{\nu_e} = - \mu_{\bar{\nu}_e}$, and $T$ is the temperature of the proto-neutron star medium (see top left panel of Fig.~\ref{fig:Fig1}). Unless otherwise specified, hereafter we will use $\hbar = c = 1$. 
In Fig.~\ref{fig:Fig1}, one can see that  the typical neutrino energies are $\mathcal{O}(100)$~MeV in the SN core and decrease  to $\mathcal{O}(10)$~MeV in the proximity of the neutrinosphere.

Another useful quantity entering our estimations is the mean-free path of the electron neutrinos and antineutrinos for neutral-current (NC) and charged-current (CC) scattering on nucleons
\begin{equation}
\label{eq:lambda}
\lambda_\nu (E, r) \simeq \sum_{\mathrm{CC, NC}} \frac{1}{\langle F_i(E,r)\rangle n_B(r) \sigma_i(E,r)} \ ,
\end{equation}
where $\sigma_i$ is the interaction cross section, $n_B$ is the number density of nucleons, and $\langle F_i(E,r)\rangle$ is the  Pauli blocking factor averaged over the neutrino energy distribution. We refer the reader to Appendix~\ref{appendix:Pauli} for a discussion on the neutrino mean free path in the SN core and for  the estimation of the Pauli blocking factor for the CC neutrino scatterings on nucleons and to  Appendix B of Ref.~\cite{Suliga:2019bsq} for the NC neutrino scattering.

\section{Sterile neutrino mixing in the supernova core}
\label{sec:conversions}
In this Section, we introduce the formalism adopted to model  flavor conversions in the $e$--$s$ sector. We discuss the neutrino mixing in matter as well as the collisional production of sterile neutrinos in the SN core.. While  the theoretical framework adopted to model the sterile neutrino production in the stellar core is outlined here,  we refer to Ref.~\cite{Suliga:2019bsq} for a detailed discussion on its numerical implementation.

\subsection{Resonant production of sterile particles}
\label{sec:MSW}

For the sake of simplicity, we work in a two-flavor basis $(\nu_e,\nu_s)$. We assume that neutrinos are Dirac particles,  sterile neutrinos do not mix with the non-electron flavors, and neglect the mixing of the active flavors among themselves.
 In fact,  flavor conversions among the active states due to the neutrino--neutrino interactions are mostly suppressed by the large matter density~\cite{EstebanPretel:2008ni}. 
Moreover, although favorable conditions for fast pairwise flavor conversions~\cite{Sawyer:2005jk,Sawyer:2015dsa,Izaguirre:2016gsx} may exist  in very small regions inside the proto-neutron star  where the $\nu_e$ degeneracy is negligible (i.e., $\mu_{\nu_e}/T\approx 0$)~\cite{Glas:2019ijo,Abbar:2019zoq,DelfanAzari:2019tez}, further numerical work is needed to investigate whether fast pairwise conversions can develop over an extended spatial region~\cite{Shalgar:2019qwg}.

For each energy mode $E$, the evolution of the neutrino (antineutrino) field is described by the Liouville equation for the density matrix  $\rho_E$ ($\bar\rho_E$)
\begin{equation}
\label{eq:denisty_matrix}
{\partial_r} \rho_E = {-i} [H_E, \rho_E] + \mathcal{C}(\rho_E,\bar\rho_E)\  \mathrm{and}\  \partial_r \bar\rho_E = {-i} [\bar{H}_E, \bar{\rho}_E] + \mathcal{C}(\rho_E,\bar\rho_E)\ ,
\end{equation}
where  $\rho_E$ is a  $2\times 2$ matrix in the flavor basis. We assume that sterile particles are generated through mixing, hence the initial conditions for the neutrino and antineutrino fields are:  $\rho_E=\mathrm{diag}(n_{\nu_e}, 0)$ and $\bar{\rho}_E=\mathrm{diag}(n_{\bar{\nu}_e}, 0)$, respectively. 
The Hamiltonian $H_E$  in the flavor basis is:
\begin{equation}
\label{hamiltonian}
H _E= H_{\mathrm{vac}, E} + H_{\mathrm{m}, E} = \frac{\Delta m_s^2}{2 E} 
\begin{bmatrix}
    -\cos 2\theta & \sin 2\theta \\
    \sin 2\theta & \cos 2\theta 
\end{bmatrix}
 + \begin{bmatrix}
    V_{\mathrm{eff}} & 0 \\
    0 & -V_{\mathrm{eff}}
\end{bmatrix}\ ,
\end{equation}
where $H_{\mathrm{vac,E}}$ describes the mixing of neutrinos in vacuum. It is a function of the active-sterile mixing angle $\theta$ and of the active-sterile mass difference $\Delta m_s^2$. We focus on  sterile neutrino rest masses in the $\mathcal{O}(1-100)~\mathrm{keV}$ range. Hence, we can approximate  $\Delta m^2_s~=~m^2_b-m^2_a \approx m^2_b \equiv m^2_s$, where  $m_b$ denotes the rest mass associated to  the extra mass eigenstate, introduced to model the mixing between active and sterile states, and $m_a$ is the  mass of one of the active eigenstates. Note that we have chosen a basis such that $H_{\mathrm{vac}}$  has opposite sign for neutrinos and antineutrinos, while $H_{m,E}$ is the same for neutrinos and antineutrinos, see e.g.~\cite{Duan:2006an}.

The term $H_{\mathrm{m,E}}$ takes into account the mixing of neutrinos in matter, and the effective potential $V_{\mathrm{eff}}$ includes the forward scattering contributions~\cite{Raffelt:2011nc}:
\begin{equation}
\label{eq:potential}
V_{\mathrm{eff}} = \sqrt{2} G_F n_B \left[\frac{3}{2}Y_e + 2 Y_{\nu_e} + Y_{\nu_\mu} + Y_{\nu_\tau} - \frac{1}{2}\right]\ ,
\end{equation}
where $G_F$ is  the Fermi constant. The fraction of particle species $i$ relative to baryons is $Y_i = (n_i - \bar{n}_i)/n_B$. 
Since $\mu_{\nu_\tau} = \mu_{\nu_\mu} \simeq 0$ and we neglect flavor conversions in the active sector, the energy distributions of the non-electron neutrinos and antineutrinos are identical and  $Y_{\nu_{\mu,\tau}}=0$. Furthermore, because of charge neutrality, $Y_p = Y_e  = 1 - Y_n$. 
\begin{figure}  
\centering
\includegraphics[scale=0.35]{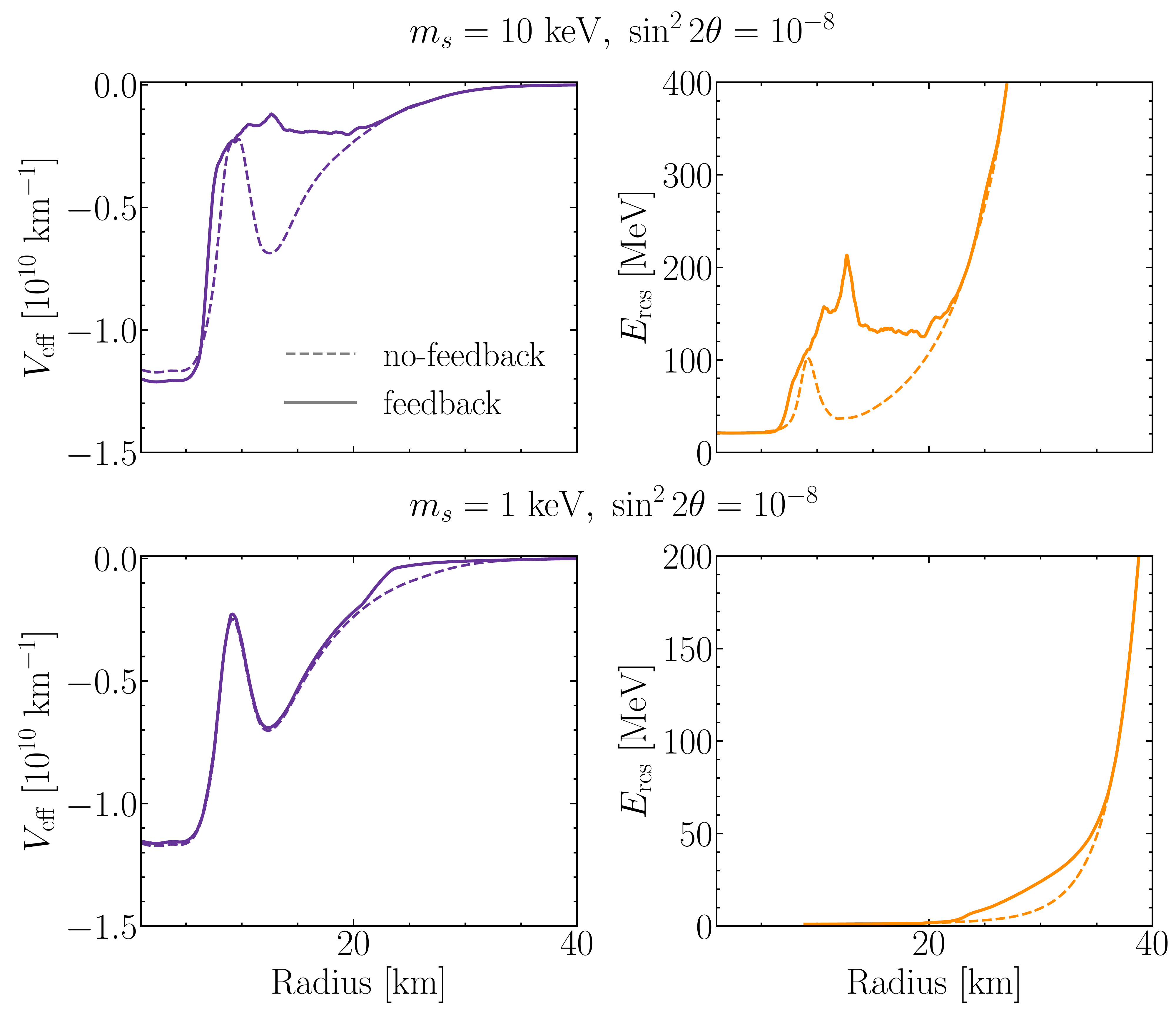}
    \caption{Initial (dashed, no-feedback, $\Delta t=0$) and final (continuous,  with dynamical feedback, $\Delta t=1$~s) radial profile of the effective matter potential (left panels) and  the resonance energy (right panels)  for $(m_s, \sin^2 2\theta)=(10~\mathrm{keV}, 10^{-8})$ on the top and $(m_s, \sin^2 2\theta)=(1~\mathrm{keV}, 10^{-8})$ on the bottom. Note that $E_{\mathrm{res}}$ is not visible below $\simeq 8$~km  in the bottom right panel, since no neutrino modes are in resonance above $1$~MeV. Because of the temporal evolution, the profile of $V_{\rm{eff}}$ is modified especially for large $m_s$.  $E_{\rm{res}}$ quickly increases as the distance from the proto-neutron star increases.}
    \label{fig:Fig2}
\end{figure}

The left panels of Fig.~\ref{fig:Fig2} show the radial profile of $V_{\rm{eff}}$ for our benchmark SN model for the case without dynamical feedback ($\Delta t=0$).  The radial profile of the effective matter potential is strongly affected from the dynamical feedback due to the sterile neutrino production ($\Delta t=1$~s), as we will discuss in Sec.~\ref{sec:Feedback}. Differently from Ref.~\cite{Hidaka:2007se} but analogously to Ref.~\cite{Warren:2014qza}, our $V_{\rm{eff}}$ is negative; hence, MSW enhanced flavor conversions are expected to occur for antineutrinos when $\mu_{\nu_e}/T \lesssim 1$ (see Fig.~\ref{fig:Fig1}, bottom left panel).

When neutrino flavor conversions occur in matter, the conversion probability is enhanced if the effective mixing angle in matter is maximal. This is the so-called  Mikheyev-Smirnov-Wolfenstein (MSW) resonance~\cite{Mikheev:1986if,1985YaFiz..42.1441M,1978PhRvD..17.2369W} and takes place when
\begin{equation}
\cos 2\theta = \frac{2 V_{\mathrm{eff}} E_{\mathrm{res}}}{m_s^2}\ .
\end{equation}
As one can see from the left panels of Fig.~\ref{fig:Fig2}, we expect multiple MSW resonances for antineutrinos~\cite{Shi:1993ee,Nunokawa:1997ct}, efficiently producing $\bar\nu_s$ when $\mu_{\nu_e}/T \lesssim 1$ (see the bottom left panel of Fig.~\ref{fig:Fig1}); on the other hand, since $V_{\rm{eff}}$ is always negative, no MSW resonances are expected for neutrinos.  The occurrence of multiple MSW resonances is a peculiar feature of the $e$--$s$ mixing, which is not found for the $\tau$--$s$ mixing due to the differences in the $V_{\mathrm{eff}}$ radial profile, see e.g.~\cite{Suliga:2019bsq}.  The multiple MSW resonances imply that $\bar\nu_e \rightarrow \bar\nu_s$, then  $\bar\nu_s \rightarrow \bar\nu_e$, and so on. As we will discuss in Sec.~\ref{sec:Feedback}, this mechanism might allow to accumulate high-energy active antineutrinos in the proximity of $R_\nu$ affecting the explosion mechanism and the proto-neutron star evolution~\cite{Shi:1993ee,Hidaka:2007se}. 

The right panels of Fig.~\ref{fig:Fig2} display the resonance energy, $E_{\mathrm{res}}$, as a function of the radius. The resonance energy $E_{\mathrm{res}}$ increases very steeply as the radius increases, therefore affecting the high-energy tail of the spectral distribution. This is in agreement with the findings of Refs.~\cite{Hidaka:2006sg,Hidaka:2007se} that  predicted a growth of $E_{\mathrm{res}}$ (for $\nu_e$ in their case and $\bar\nu_e$ in our case) towards the outer edge of the proto-neutron star within a one-zone model. 
As we will discuss in Sec.~\ref{sec:Feedback}, the dynamical feedback due to the production of sterile particles dramatically affects $V_{\mathrm{eff}}$ and, as a consequence, $E_{\mathrm{res}}$, especially for large mixings. Note that $E_{\mathrm{res}}$  is not visible for $ r \lesssim 8$~km  in the bottom right panel; this is because no neutrino modes are in resonance above $1$~MeV, where the latter has been assumed as minimum energy in our numerical simulations.

\begin{figure}
\centering
\includegraphics[scale=0.4]{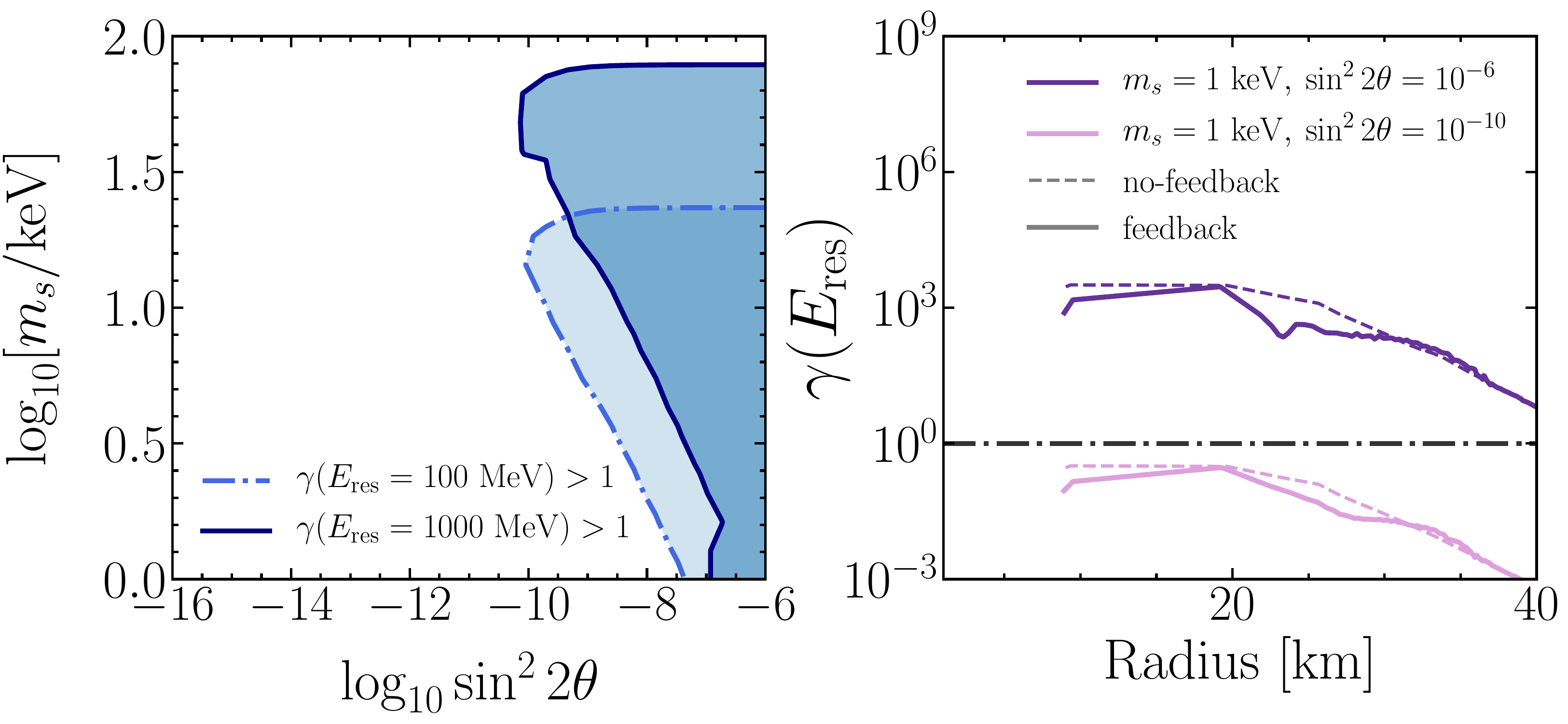}
\caption{{\it Left:} The shaded regions highlight the $(\sin^2 2\theta, m_s)$ mixing parameters such that  the adiabaticity parameter $\gamma$ is greater or equal to unity for antineutrinos with $E=100$~MeV (light blue) and $E=1000$~MeV (dark blue) in the absence of dynamical feedback due to the production of sterile particles.  {\it Right:} Radial profile of the adiabaticity parameter at the resonance energy without dynamical feedback (dashed lines) and with feedback (solid lines). The violet [pink] lines show the results for $(m_s,\sin^2 2\theta)=(1~\mathrm{keV}, 10^{-6})$ [$(m_s,\sin^2 2\theta)=(1~\mathrm{keV}, 10^{-10})$]. The flavor conversions are adiabatic only for very large $\sin^2 2\theta$ and large $m_s$. The adiabaticity parameter, $\gamma$, is not strongly affected by the dynamical feedback.} 
\label{fig:adiabaticity}
\end{figure}
The MSW resonant conversion can take place only when $\lambda_\nu$ is larger than the resonance width:
\begin{equation}
\Delta_{\mathrm{res}} = \tan 2\theta \left|\frac{dV_{\mathrm{eff}}/dr}{V_{\mathrm{eff}}}\right|^{-1}\ . 
\end{equation} 
If this condition is satisfied, the Landau-Zener formula~\cite{Kim:1987ss,Parke:1986jy} describes well the electron-sterile conversion probability 
\begin{equation}
\label{eq:P_msw}
P_{\rm{e s}}(E_{\mathrm{res}}) = 1- \exp\left({-\frac{\pi^2}{2}\gamma}\right)\ ,
\end{equation}
with $\gamma = \Delta_{\mathrm{res}}/l_{\mathrm{osc}}$ being the adiabaticity parameter; the oscillation length at resonance is $l_{\mathrm{osc}} = (2 \pi E_{\mathrm{res}})/( m_s^2 \sin 2\theta)$. 
In order to investigate the adiabaticity of MSW conversions, the shaded regions in the left panels of Fig.~\ref{fig:adiabaticity} highlight when $\gamma \ge 1$ in the $(\sin ^2 2\theta, m_s)$ parameter space at the radius where  the antineutrino energy modes $E = 100$ and $1000$~MeV are in resonance. One can see that the assumption of full adiabaticity of flavor conversions often adopted in the literature for simplicity, see e.g.~Ref.~\cite{Warren:2014qza}, only holds for  $\sin^2 2\theta \gtrsim 10^{-7}$. The dark blue region shows a little kink for $m_s \simeq 1$~keV because the  maximum antineutrino energy that goes into resonance before $R_\nu$ is smaller than $1000$~MeV. The radial profile of $\gamma$ at the resonant energy $E_{\mathrm{res}}$ shown in the right panel of  Fig.~\ref{fig:adiabaticity} suggests that MSW-enhanced flavor conversions are adiabatic for large mixings and flavor conversions become less adiabatic for smaller mixing angles. Interestingly, the dynamical feedback due to the production of sterile particles (see Sec.~\ref{sec:Feedback}) only affects $\gamma$ insignificantly,
as evident from the comparison between the dashed and the continous line in the right panel of Fig.~\ref{fig:adiabaticity}.

\subsection{Collisional production of  sterile particles}
\label{sec:coll_production}

The dense SN core is an ideal environment for the sterile neutrino production by scattering-induced decoherence~\cite{Raffelt:1992bs}.   
In fact, due to collisions on nucleons, each propagation eigenstate collapses into a flavor eigenstate possibly 
leading to the prodution of sterile particles. 
The  in-medium conversion probability  taking into account the collisional production is
\begin{equation}
\label{eq:Pxs_collisions}
 \langle P_{\mathrm{e s}}(E, r) \rangle \approx \frac{1}{4} \frac{\sin^2 2\theta}{(\cos 2\theta - 2V_{\mathrm{eff}}E/m_s^2)^2 +\sin2\theta^2 + D^2}\ ,
\end{equation}
with $D = [E \Gamma_{\nu}(E)]/{m_s^2}$ being the quantum damping term.

Electron neutrinos and antineutrinos undergo NC and CC scatterings with a total  collision rate 
\begin{equation}
\label{eq:coll_rate}
\Gamma_\nu (E) = n_B \left[\langle F_{\mathrm{NC}, \mathrm{p}}(E) \rangle Y_e \sigma_{\nu \mathrm{p, NC}}(E) + \langle F_{\mathrm{NC}, \mathrm{n}}(E)\rangle (1 - Y_e) \sigma_{\nu \mathrm{n, NC}}(E) + \langle F_{\mathrm{CC}, \mathrm{N}}(E)\rangle Y_\mathrm{N} \sigma_{\nu \mathrm{N, CC}}(E)  \right]\ ,
\end{equation}
where $\langle F_{\mathrm{NC, CC}, \mathrm{p, n}}(E)\rangle$ represents the Pauli blocking factor averaged over the neutrino energy spectrum and computed as in Appendix~\ref{appendix:Pauli} 
(see also Appendix~A of Ref.~\cite{Suliga:2019bsq}).
The label $N$ in $\Gamma_\nu (E)$ in the CC contribution  depends on whether the rate is calculated for the electron neutrinos ($\mathrm{N = n}$) or antineutrinos ($\mathrm{N = p}$).  Note that, except for the correction due to collisions, Eq.~\ref{eq:Pxs_collisions} is defined as the effective conversion probability of neutrinos in matter.

To compute the sterile neutrino conversions and reconversions self-consistently, we use the effective conversion probability as from Appendix~B of Ref.~\cite{Suliga:2019bsq}: 
\begin{equation}
\label{eq:P_coll}
P_{\rm{es}} (E, n) = \frac{1}{n}\sum_{k=1}^{n}   \langle P_{\rm{es}} \rangle (1 - \sin^2 2\widetilde\theta \langle P_{\rm{es}}\rangle )^{k - 1}\ ,
\end{equation}
where $\widetilde\theta$ is the effective mixing angle in matter under the assumption that the volume within the neutrinosphere is divided in many sub-shells where the matter potential is  constant; the number of sub-shells within a SN region of width $\Delta r$ is $n =  \left[\Delta r/\lambda_\nu\right]$.

\subsection{Characterization of the production of sterile particles}
\label{sec:sterile_neutrino_production}
\begin{figure}[t] 
\centering
\includegraphics[scale=0.34]{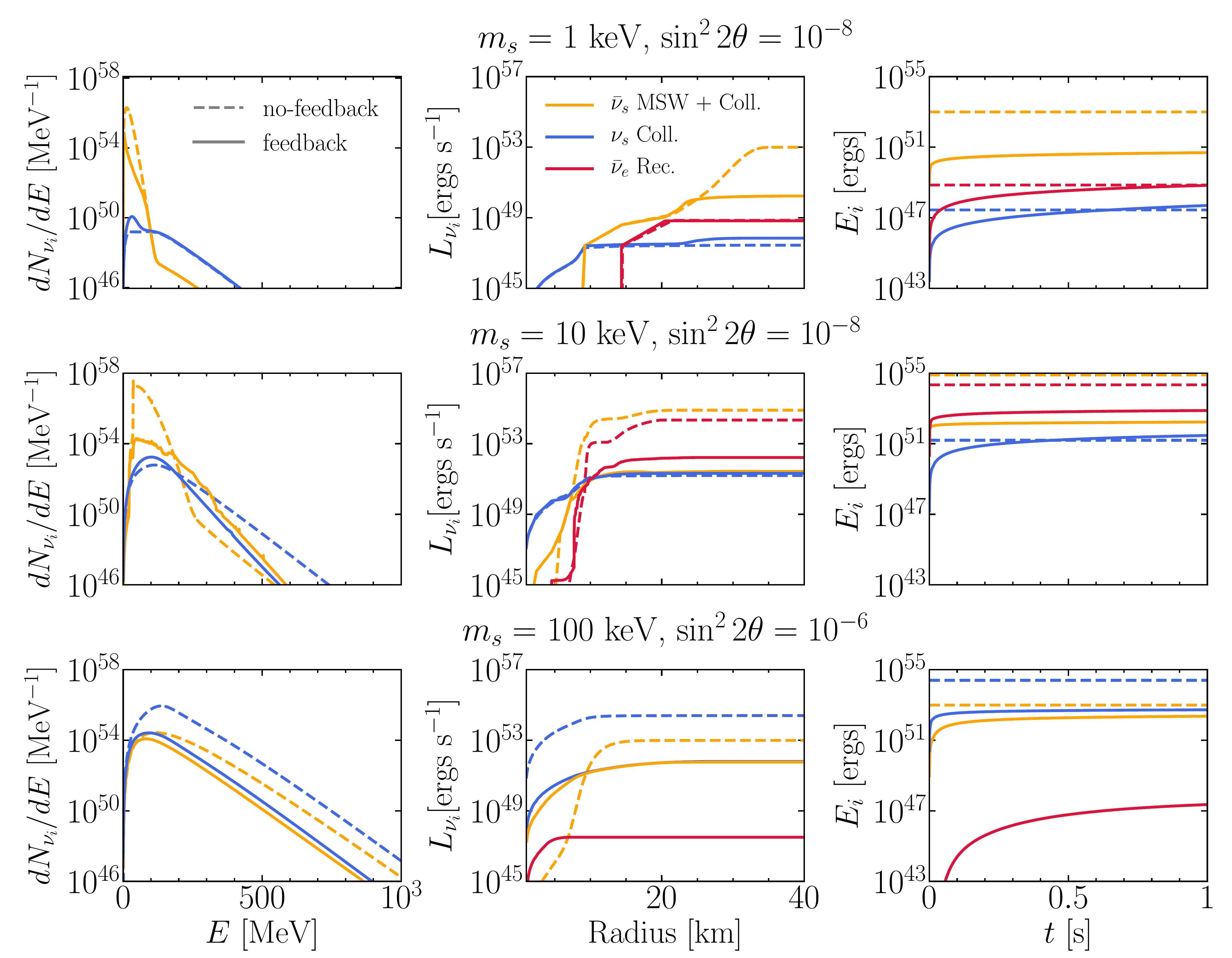}
    \caption{{\it Left:} Energy distribution of  sterile neutrinos (blue) and antineutrinos (orange). {\it Middle:} Cumulative luminosity as a function of the radius. {\it Right:} Energy emitted in the sterile neutrinos (blue) and sterile antineutrinos (orange), and energy reconverted back to the active antineutrinos (red) within $R_\nu$ as a function of time.  On all panels the dashed lines represent the calculations without including the dynamical feedback effects due to the production of sterile particles, whereas the solid lines include the feedback effects. The top panels have been obtained for $(m_s, \sin^2 2\theta)=(1~\mathrm{keV}, 10^{-8})$, the middle ones for $(m_s, \sin^2 2\theta)=(10~\mathrm{keV}, 10^{-8})$, and the bottom ones for $(m_s, \sin^2 2\theta)=(100~\mathrm{keV}, 10^{-6})$.  To highlight  the fact  that  no time evolution is computed in the case without feedback, $E_i$ is plotted as a constant.
    Sterile neutrinos are mainly produced collisionally, while sterile antineutrinos are produced through MSW conversions and collisions; sterile antineutrinos are therefore more abundant for sterile mixing parameters favoring  MSW conversions. The dynamical feedback has a non-negligible contribution on the production of sterile particles.
    }
    \label{fig:steriles}
\end{figure}
In our numerical implementation, since neutrinos and antineutrinos are trapped within the neutrinosphere, we compute the number of $\nu_s$'s and $\bar\nu_s$'s at each radius by taking into account the various production mechanisms described in Secs.~\ref{sec:MSW} and \ref{sec:coll_production} (MSW conversions, collisional production, and reconversions), each with its own efficiency.
The left panels of Fig.~\ref{fig:steriles} illustrate the energy distributions of sterile neutrinos and antineutrinos as a function of the neutrino energy at the neutrinosphere in the absence of dynamical feedback (dashed lines) and with dynamical feedback (continuous lines, see Sec.~\ref{sec:Feedback}), for three different mixings: $(m_s, \sin^2 2\theta)=(1~\mathrm{keV}, 10^{-8})$,  $(m_s, \sin^2 2\theta)=(10~\mathrm{keV}, 10^{-8})$, and  $(m_s, \sin^2 2\theta)=(100~\mathrm{keV}, 10^{-6})$ from top to bottom, respectively. Our findings are in agreement with the ones in Ref.~\cite{Suliga:2019bsq}:
since sterile neutrinos are mostly produced through collisions, while sterile antineutrinos are resonantly and collisionally produced, sterile antineutrinos are more abundant than neutrinos. 
As discussed in Sec.~\ref{sec:MSW} and in Ref.~\cite{Suliga:2019bsq}, the production of sterile particles is less abundant for smaller $m_s$ because MSW conversions occur for these masses at larger distances from the core where the number density of active neutrinos is lower.

The energy distribution of sterile neutrinos resonantly and collisionaly produced within the neutrinosphere after a time $\Delta t$ is~\cite{Suliga:2019bsq} 
\begin{gather}
\label{eq:d_nu_dE}
\begin{aligned}
\frac{d \mathcal{N}_{\nu_s}}{dE} (t_\mathrm{pb}+\Delta t) ={}& \sum_{l=1}^{P} \sum_{i=1}^{N} \left[ {P}_{\rm{es}}(E, r_i, t_l) \frac{dn_{\nu_e}}{dE}(r_i, t_l) 
 - {P}_{\rm{se}}(E, r_i, t_l) \sum_{j=1}^{i-1}  \frac{dn_{\nu_s}}{dE}(r_j, t_l)  \frac{r_j^2}{r_i^2} \right] \\ & \times \ \Delta t_l  \  \Delta V_i \ {\Delta r_ i}^{-1} \ ,
\end{aligned}
\end{gather}
where  $[t_1, t_P ] = [t_\mathrm{pb}, t_\mathrm{pb} + \Delta t]$ and $[r_1, r_N] = [1~\mathrm{km}, R_\nu]$.  The size of each radial step is $\Delta r_i \equiv r(E_i ) - r(E_{i-1})$, where $E_i$ and $E_{i-1}$ are the resonance energies, and the differential energy-dependent volume of each SN shell is $\Delta V_i = 4 \pi r_i^2 \Delta r_i$  and the radius on which the SN shell is centered is $r_i$. $P_{\rm{es}}$ $(\bar{P}_{\rm{es}})$ is the $e$--$s$ neutrino (antineutrino) conversion probability given by Eq.~\ref{eq:P_msw} 
for the MSW conversions and Eq.~\ref{eq:P_coll} for the collisional production. $dn_{\nu_e} / dE$ is the local density of active neutrinos in the SN shell of width $\Delta r_ i$, per unit energy. 
The first term in Eq.~\ref{eq:d_nu_dE} accounts for the conversions from active to sterile states, while the second term in the above equation represents the reconversion of sterile particles to the electron flavor, where $P_{\rm{se}} (\bar{P}_{\rm{se}})$ is the sterile-electron  reconversion probability; the latter is assumed to be non-zero and equal to $P_{\rm{es}} (\bar{P}_{\rm{es}})$ when the resonance condition is satisfied, while the term $r_j^2/r_i^2$ takes into account the dilution of neutrinos as they approach the outer layers of the proto-neutron star. The sterile neutrino density per energy unit  is
${dn_{\nu_s}}/{dE}  = P_{\rm{es}} \ {dn_{\nu_e}}/{dE}$,
and  the inner sum in Eq.~\ref{eq:d_nu_dE} spans the radial range $[1~\mathrm{km}, r_{i-1}]$. An analogous expression holds for antineutrinos.

The cumulative sterile neutrino luminosity at each radius $r$ inside the neutrinosphere is 
\begin{gather}
\label{eq:Lum_nu_s}
\begin{aligned}
L_{\nu_s} (r, t_\mathrm{pb}+\Delta t) ={}& \sum_{i=1}^{i} \sum_{k=1}^{L} \left[ {P}_{\rm{es}}(E_k, r_{i}, t_l) \frac{dn_{\nu_e}}{dE}(r_i, t_l)
 - {P}_{\rm{se}}(E_k, r_{i}, t_l) \sum_{j=1}^{i-1}  \frac{dn_{\nu_s}}{dE}(r_j, t_l)  \frac{r_j^2}{r_{i}^2} \right] \\ & \times \ 4\pi r_{i}^2  \ \Delta E_k \ E_k \ \Delta V_i \ {\Delta r_i}^{-1}\ ,
\end{aligned}
\end{gather}
where $[E_1, E_L] = [1, 10^3]~\mathrm{MeV}$ and $\Delta E_k = 1~\mathrm{MeV}$.
The cumulative luminosity is shown in the middle panels of Fig.~\ref{fig:steriles}, one can see that most of the production of sterile particles occurs within  $20$~km from the SN core. 
For the $m_s =100$~keV case, no reconversions occur when the dynamical feedback is not included, as the energy modes undergoing MSW resonances are larger than $1000$~MeV.

The total energy going  into sterile particles within $R_\nu$ in $\Delta t= 1$~s  is:
\begin{equation}
\label{eq:Es}
E_{\nu_s} = \sum_{l = 1}^{P} L_{\nu_s}(R_{\nu}, t_l) \Delta t_l \ ,
\end{equation}
with  $\Delta t _l = 10^{-4}~\mathrm{s}$; we assume that  all sterile particles stream freely from
the neutrinosphere every $\Delta t_l$ to take into account the reconversion effects.
The right panels of Fig.~\ref{fig:Fig2} show the energy emitted in sterile neutrinos and  antineutrinos  (with dashed lines for the case without dynamical feedback) and the energy reconverted back from sterile to active antineutrinos. In agreement with what observed for the energy distributions (left panels), the amount of energy emitted in sterile antineutrinos is larger than the one emitted in sterile neutrinos and the overall sterile energy budget due to MSW conversions and collisions increases with $m_s$. The reconversions are most efficient for intermediate masses $m_s \sim 10$~keV (middle panels); the reason being that the MSW region occurs around the  minimum of $V_\mathrm{eff}$ for the energy modes close to the peak of the neutrino energy distribution, as it can be seen on the right upper panel of Fig~\ref{fig:Fig2}. For high masses, $m_s \gtrsim 50$~keV (see bottom panels),  initially there are no resonances for $E \lesssim 1000~\mathrm{MeV}$. On the other hand, the MSW region for the small masses $m_s \simeq 1~\mathrm{keV}$ (top panels) starts relatively far away from the SN center (after the potential minimum) and this suppresses the efficient production of sterile particles through collisions. Notably, we do not observe any high-frequency modulation of the luminosity as found in Ref.~\cite{Warren:2014qza}; this may be due to the fact that we adopt  high temporal and spatial resolution that allows for convergence, smoothing any numerical instability, or because of the differences in the modeling of the neutrino flavor conversions.

As we will discuss in detail in Sec.~\ref{sec:Feedback}, most of the production of sterile particles occurs within the first $0.4$~s for our SN model when the dynamical feedback is included.  The $m_s = 1$~keV case (top panels) reaches the saturation energy more slowly than the  $m_s = 10$~keV case (middle panel) because the MSW region is further away from the temperature maximum for the $m_s = 1$~keV case; hence,  it takes longer time for $Y_s$ to develop.

\newpage
\section{Dynamical feedback due to the production of sterile particles}
\label{sec:Feedback}

In this Section, we describe
the method adopted to implement the feedback on flavor  conversions into sterile states  inside the proto-neutron star. After introducing  the evolution equations adopted for implementing the dynamical feedback, we investigate the impact of the production of sterile particles  on the chemical potentials of electrons, neutrinos, protons and neutrons, the medium temperature, the entropy, and the propagation of the sterile states themselves. 

\subsection{Dynamical feedback equations}
\label{sec:equations}
\subsubsection{Evolution of the chemical potentials and particle fractions}
\label{sec:Beta}

Due to the very high density and temperature inside the proto-neutron star core, the $\beta$-equilibrium is sustained  for the following CC reactions: 
\begin{equation}
\label{eq:beta_reac}
\mathrm{e^-} + \mathrm{p} \leftrightarrow \nu_e + \mathrm{n} \quad \mathrm{and} \quad \mathrm{e^+} + \mathrm{n} \leftrightarrow \bar{\nu}_e + \mathrm{p} \ .
\end{equation}

Since the production of sterile (anti)neutrinos affects the $\nu_e$ lepton number, this in turn affects the composition of electrons and positrons, as well as the neutrons and protons through $\beta$-equilibrium.
We take into account the feedback effect due to the production of sterile states, by solving the following system of  coupled equations as functions of the distance from the SN core, $r$, and time $t$
\begin{align}
 &   \mu_e(r, t) + \mu_{\mathrm{p}}(r, t) + m_{\mathrm{p}} =  \mu_{\nu_e}(r, t) + \mu_{\mathrm{n}}(r, t) + m_{\mathrm{n}} \label{eq:coupled_1} \ ,\\
  &  Y_\mathrm{e}(r, t) + Y_{\nu_e}(r, t) + Y_{\nu_s}(r, t) = {\rm{const.}} \label{eq:coupled_2} \ , \\
   & Y_\mathrm{p}(r, t) + Y_\mathrm{n}(r, t) =  1 \label{eq:coupled_3} \ ,\\
 &   Y_\mathrm{p}(r, t) =  Y_e(r, t)  \label{eq:coupled_4}  \ ,
\end{align}
where $m_{p}$ ($m_n$) is the proton (neutron) mass; i.e.~we impose the conservation  of the chemical potential in the $\beta$-equilibrium reactions,  the lepton number conservation,  
 the charge conservation, and  the baryon number conservation.  
  
 The sterile neutrino lepton number at the time $t$ and at any specific radius $r$ is
\begin{gather}
\label{eq:Y_nu_s}
\begin{aligned}
Y_{\nu_s}(r, t) ={}& 
\frac{1}{n_B(r)} \sum_{l=1}^{P} \sum_{k=1}^{L} 
\left[ 
{P}_{\rm{es}}(r, E_k, t_l) \frac{dn_{\nu_e}}{dE_k}(r, t_l) - 
\bar{P}_{\rm{es}}(r, E_k, t_l) \frac{dn_{\bar{\nu}_e}}{dE_k}(r, t_l) \right. \\ &
\left. - P_{\rm{se}}(r, E_k, t_l) \sum_{j=1}^i P_{\rm{es}} (r_j, E_k, t_l)  \frac{dn_{{\nu}_e}}{dE_k}(r_j, t_l) \frac{r_j^2}{r_i^2}  \right. \\ &
 \left. + \bar{P}_{\rm{se}}(r, E_k, t_l) \sum_{j=1}^i \bar{P}_{\rm{es}} (r_j, E_k, t_l) \frac{dn_{\bar{\nu}_e}}{dE_k}(r_j, t_l) \frac{r_j^2}{r_i^2}  \right]  \times \Delta E_k \ \Delta r^{-1} \ \Delta t_l \ ,
\end{aligned}
\end{gather}
where $r_i \equiv r$.

\subsubsection{Evolution of the supernova core temperature}
\label{sec:Heating}

When solving Eqs.~\ref{eq:coupled_1}--\ref{eq:coupled_4}, the thermodynamic properties of matter, $\rho(r,t)$ and $T(r,t)$ need to be provided.
Because of the shape of $V_{\mathrm{eff}}$, the sterile particles produced deep inside in the proto-neutron star core can be reconverted
into active neutrinos in the outer layers of the core, see Sec.~\ref{sec:MSW}. Since the antineutrinos undergoing MSW resonances in the SN have high energies, when they are reconverted 
from sterile states to active ones in the outer layers of the proto-neutron star,  they can  introduce an additional source of  heating for  the medium. 
This serves as an additional energy transfer  that can cool the inner part of the proto-neutron star while heating up the outer layers.
Consequently, the temperature profile of the proto-neutron star can be affected significantly.

To model this effect, we keep track of the entropy change in a fluid element:
\begin{equation}
\label{entropy}
S(r, t) = S(r, t - \Delta t_l) + \Delta S(r, \Delta t_l) \ ,
\end{equation}
with the change of entropy in $\Delta t_l$ being
\begin{equation}
\label{eq:Delta_entropy}
\Delta S(r, \Delta t_l) = \frac{ \left( Q^{h}_{\nu_s} (r, t - \Delta t_l)  -  Q^{c}_{\nu_s} (r, t -\Delta t_l) + \mu_{\nu_e} (r, t - \Delta t_l) \Delta {Y}_{\nu_s} (r, t - \Delta t_l)  \right) \Delta t_l }{T(r, t - \Delta t_l)} \ ,
\end{equation}
where $Q^{h,c}_{\nu_s}$ are the additional heating and cooling rates due to the sterile neutrino production and reconversion described in  Appendix~\ref{sec:app_qcdot}. 
This equation comes from the first law of thermodynamics,
$dS= Q/T + P/T dV - \sum_i \mu_i/T dY_i$,
by assuming that the changes in $dV$ can be neglected.
For a given EoS, the new medium temperature $T$ at the time $t \equiv t_\mathrm{pb} + \Delta t_l$ has been computed by using $S$ and $Y_e$ at a fixed $\rho$ as detailed in  Appendix~\ref{appendix:Temperature} and by employing the CompOSE package~\cite{Typel:2013rza}.

\subsection{Dynamical feedback on the electron fraction and chemical potentials}
\label{sec:Ye_mu}
\begin{figure}[ht]
\centering
\includegraphics[scale=0.32]{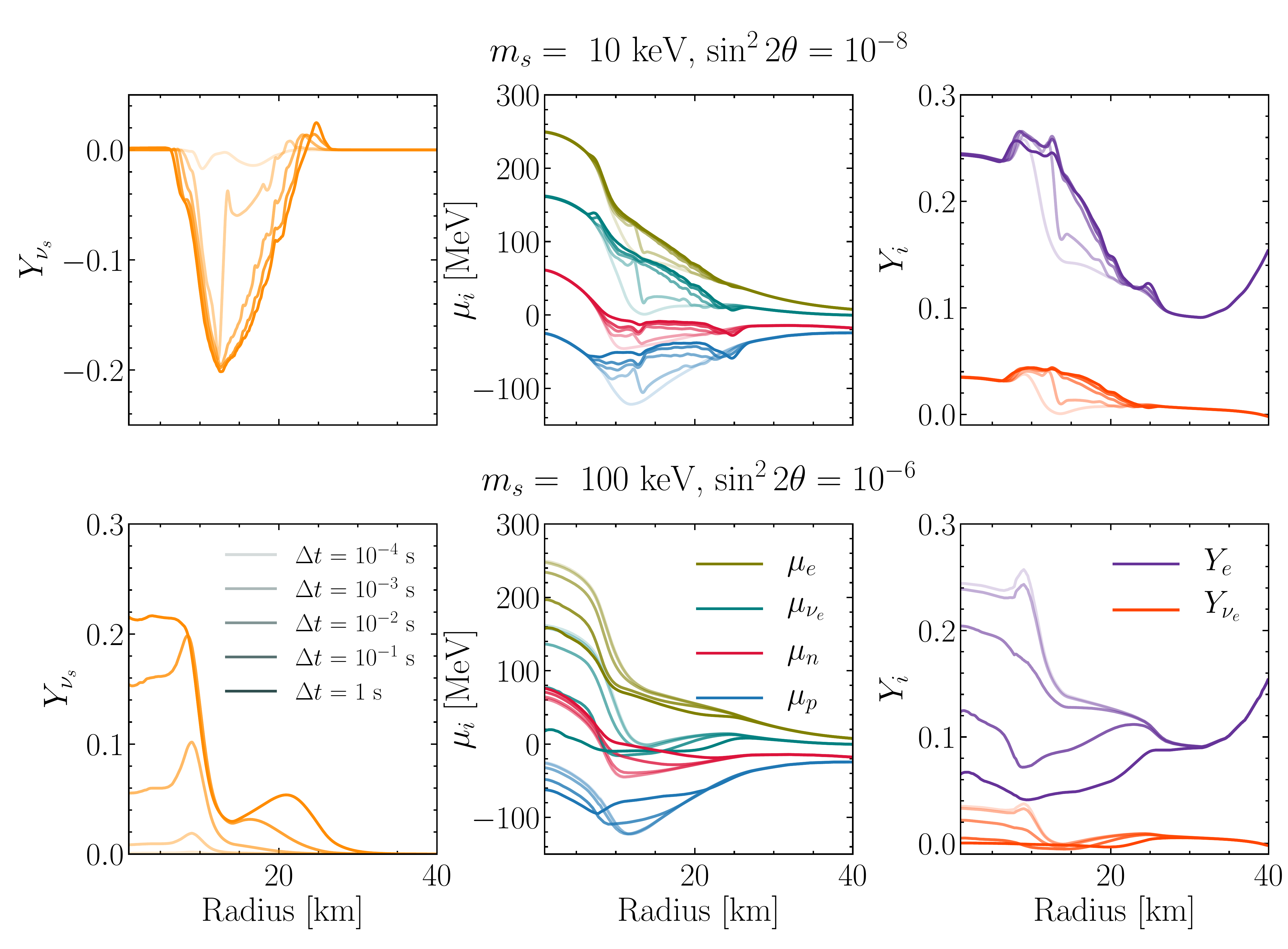}
\caption{Radial profile of the sterile neutrino fraction in the left panels, chemical potentials for electron neutrinos (teal), electrons (olive green), protons (blue), and neutrons (red) in the middle panels, and electron (purple) and electron neutrino (orange) fractions in the right panels. The temporal evolution of each of these quantities is shown for $(m_s, \sin^2 2\theta)= (10~\mathrm{keV}, 10^{-8})$ on the top panels and for $(m_s, \sin^2 2\theta)= (100~\mathrm{keV}, 10^{-6})$ on the bottom panels. The time evolution is represented by curves with different hues of the same color, from lighter to darker as $\Delta t$ increases. Because of the dynamical feedback, $Y_{\nu_s}$ decreases [increases] with time, while the $\mu_{e,{\nu_e}}$ chemical potentials increase [decrease] and $Y_i$ increase [decrease] with time for $(m_s, \sin^2 2\theta)= (10~\mathrm{keV}, 10^{-8})$ [$(m_s, \sin^2 2\theta)= (100~\mathrm{keV}, 10^{-6})$]. The sign of $Y_{\nu_s}$ changes from negative to positive as the sterile neutrino mass increases because the number of $\bar\nu_e$ energy modes undergoing resonantly enhanced conversions  decreases.} 
\label{fig:feedback}
\end{figure}
We now investigate the role that the dynamical feedback, modeled as described in Sec.~\ref{sec:equations}, has on the electron fraction and the chemical potentials of particles.
Figure~\ref{fig:feedback} shows the radial profile of the the sterile neutrino fraction,  the chemical potentials, as well as the electron and electron neutrino fractions, from left to right respectively. The temporal evolution of each of the quantities above is  represented by curves with different hues of the same color, from lighter to darker as $\Delta t$ increases.  As 
discussed for Figs.~\ref{fig:Fig2} and \ref{fig:steriles}, the radial range where flavor conversions are most efficient depends on  $m_s$; the largest feedback on the particle fractions and chemical potentials occurs in the same radial region.

Due to the initial shape of  $V_{\rm{eff}}$ (see Fig.~\ref{fig:Fig2}), only electron antineutrinos have matter-enhanced conversion probabilities. This leads to the development of a negative sterile neutrino fraction $Y_{\nu_s}$ and to an increase of the electron and electron neutrino fractions for $(m_s, \sin^2 2\theta)= (10~\mathrm{keV}, 10^{-8})$, as shown in the top panels of Fig.~\ref{fig:feedback}, as a result of the lepton number conservation (Eq.~\ref{eq:coupled_2}). As a consequence of $Y_{\nu_s}$ becoming increasingly  negative, the chemical potentials  of  electrons and  electron neutrinos increase as a function of time, see Eq.~\ref{eq:coupled_1}.
For the $(m_s, \sin^2 2\theta)= (100~\mathrm{keV}, 10^{-6})$ case, plotted in the bottom panels of Fig.~\ref{fig:feedback}, an opposite trend holds for $\mu_{\nu_e}$ and $\mu_e$, as $Y_{\nu_s}$ increases; $\mu_{n,p}$ do not monotonically decrease as a function of time because of the medium temperature effects.   We refrain from showing the $(m_s, \sin^2 2\theta)= (1~\mathrm{keV}, 10^{-8})$ case (see Fig.~\ref{fig:steriles}) because the radial and temporal evolution of the chemical potentials, and particle fractions closely follow the  trend shown for $(m_s, \sin^2 2\theta)= (10~\mathrm{keV}, 10^{-8})$.

For $(m_s, \sin^2 2\theta)= (10~\mathrm{keV}, 10^{-8})$, the feedback effects are negligible at larger radii, in the proximity of $R_\nu$, because the MSW resonance energies are very large in that region  (as it can be seen from the upper right panel of Fig.~\ref{fig:Fig2}); hence, only very few electron antineutrinos with  large energies may undergo resonances. For both mixings, the production of sterile neutrinos does not affect the particle fractions at higher radii as well as the chemical potentials. This is because, in addition to the inefficient MSW production, collisions are inefficient in producing high numbers of sterile neutrinos and antineutrinos due to the declining temperature; in fact,  the number of sterile antineutrinos produced deep inside the SN core, through collisions and MSW resonances, is smaller than  the number of sterile antineutrinos produced after the temperature maximum (see Fig.~\ref{fig:Fig1}). 

Since the effective matter potential felt by neutrinos depends on $Y_e$ and $Y_{\nu_e}$ (Eq.~\ref{eq:potential}), $V_{\rm{eff}}$ is also affected, as shown in Fig.~\ref{fig:Fig2}. 
 Similarly to what was found for the $\tau$-$s$ mixing in Ref.~\cite{Suliga:2019bsq}, Fig.~\ref{fig:steriles} shows that the total energy going into sterile states as a function of time is much smaller than the one obtained in the case without feedback.
For $(m_s, \sin^2 2\theta)= (10~\mathrm{keV}, 10^{-8})$, the energy of the modes undergoing MSW resonances is higher as $\Delta t$ increases because $V_{\mathrm{eff}}$ rises as a consequence of the dynamical feedback (see Fig.~\ref{fig:Fig2}). This affects the resonance energy $E_{\rm{res}}$ as is displayed in Fig.~\ref{fig:Fig2} and in Fig.~\ref{fig:steriles} for the $m_s =10$~keV case  where resonantly-enhanced conversions occur. At the same time, given that a smaller number of energy modes undergo resonances as $\Delta t$ increases, the sterile antineutrino production declines over time. 
Additionally, the production of sterile antineutrinos is reduced with respect to the case without feedback (see Fig.~\ref{fig:steriles}) because the dip in $V_{\mathrm{eff}}$ appearing in correspondence of the temperature maximum for $\Delta t=0$, disappears as $\Delta t$ increases. The sterile  antineutrinos produced deep inside the core, e.g.~below $\sim$~15~km, are not abundant because of the high electron neutrino degeneracy (large $\mu_{\nu_e}$). The adiabaticity of the resonantly enhanced production of sterile states is, however, not affected as shown in Fig.~\ref{fig:adiabaticity}.

For $(m_s, \sin^2 2\theta) = (100~\mathrm{keV}, 10^{-6})$, initially the energy modes below $1000$~MeV are prevented to undergo MSW resonances, and the  neutrino degeneracy is large, hence the collisonal production of sterile neutrinos dominates over antinuetrinos, as can be seen from  Fig.~\ref{fig:steriles}. Consequently, this leads to a decrease of $Y_e$ and $Y_{\nu_e}$ and the corresponding chemical potentials (because we impose the lepton number conservation, Eq.~\ref{eq:coupled_2}). As $\Delta t$ increases, this trend causes  $V_\mathrm{eff}$ to decrease and allows  antineutrinos with energies $\lesssim 1000~\mathrm{MeV}$ to undergo resonantly enhanced conversions and reconversions (this explains the appearance of the red line for the feedback case in Fig~\ref{fig:steriles}). The decreasing chemical potential for electron neutrinos is responsible for nearly equilibrating the produced sterile neutrinos and antineutrinos energy distributions (see Fig~\ref{fig:steriles}).

\subsection{Implications of the dynamical feedback on the electron fraction}
\label{sec:Ye}

The electron fraction is an important quantity regulating the physics of the SN core.  
During the infall stage, a higher $Y_e$ may be responsible for a larger  bounce-shock energy, altering the explosion mechanism as discussed e.g.~in  Ref.~\cite{Fuller:1981mu}.
On the other hand, during the accretion phase, 
 a higher deleptonization rate leading to a lower $Y_e$ could result in larger neutrino luminosities of all flavors and enhance the neutrino heating behind the SN shock~\cite{Warren:2014qza}.

\begin{figure}[t]
\centering
\includegraphics[scale=0.4]{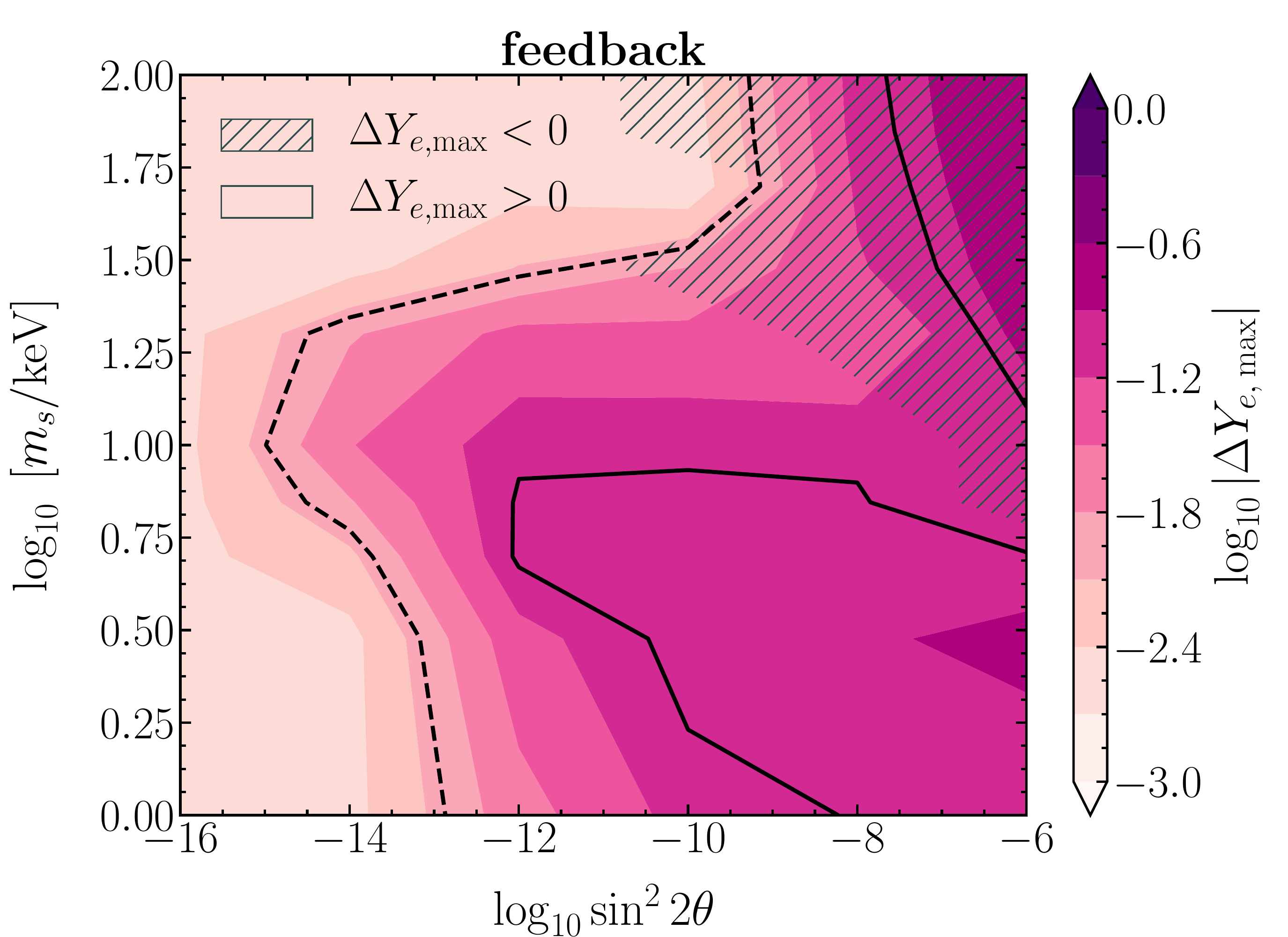}
\caption{Contour plot of the maximal difference between the final ($\Delta t = 1~\mathrm{s}$, with dynamical feedback) and the initial ($\Delta t = 0~\mathrm{s}$)  electron fraction $Y_e$ in the ($\sin^2 2\theta$, $m_s$) parameter space.
 Depending on the mixing parameters, the production of sterile particles can induce an increase (non-hatched region) or a decrease (hatched region) of $Y_e$ with respect to the case without feedback. A maximum $|\Delta Y_e| \simeq 0.21$ is achieved in the purple region for large $\sin^2 2\theta$.} 
\label{fig:contour_ye}
\end{figure}
In order to quantify the variation of $Y_e$ due to the sterile neutrino production, as a function of ($m_s, \sin^2 2\theta$), Fig.~\ref{fig:contour_ye} displays the maximal decrease or increase of $Y_e$ within the neutrinosphere with respect to its initial value: $\Delta Y_e = Y_{e, \mathrm{final}} - Y_{e, \mathrm{initial}}$. The maximal variation of $|\Delta Y_e|$ is $\simeq 0.21$  for large $\sin^2 2\theta$. As shown in Fig.~\ref{fig:feedback}, $Y_e$  increases or decreases with respect to its original value according to the oscillation physics.
For large values of $m_s$ and large $\sin^2 2\theta$, only very high-energy modes undergo MSW resonances; hence, the production of sterile particles is initially mostly driven by collisions. Since collisions tend to equilibrate the spectral distributions of $\nu_e$ and $\bar\nu_e$,  $Y_e$ decreases. For smaller values of $m_s$ ($m_s \lesssim 50~\mathrm{keV}$) and large $\sin^2 2\theta$, the minimum of $V_{\mathrm{eff}}$ becomes increasingly negative and a larger fraction of the energy spectrum of $\bar\nu_e$'s  undergoes MSW transitions. Since only antineutrinos are affected by MSW resonances,  this implies an increase of $Y_e$ with time. In all cases,  as $\Delta t$ increases, the number of MSW resonances affecting each energy mode decreases (see Fig~\ref{fig:Fig2}). Intermediate sterile masses,  $m_s \simeq 10$--$30$~keV, have a variable  sign for $\Delta Y_e $. For small mixing angles,  $Y_e$ increases,  but for large $\sin^2 2\theta$ the collisional production of $\nu_s$ (and $\bar \nu_s$ although in a smaller fraction that  increases because of  reconversions) is larger; as a result, $Y_e$ decreases in the inner SN core and increases in the outer part of the SN core.

Importantly, as shown in Fig.~\ref{fig:feedback}, the variation of $Y_e$ occurs within the proto-neutron star and we do not find any impact due to the dynamical feedback in the spatial region  closely surrounding the neutrinosphere. As a consequence, we expect that the $Y_e$ variation  illustrated in Fig.~\ref{fig:contour_ye} may contribute to affect the evolution of the proto-neutron star and may only indirectly impact the SN explosion mechanism by modifying the neutrino emission properties. Our findings are in general agreement with the ones  of  Ref.~\cite{Warren:2014qza}, but are quantitatively different because of the different $(m_s,\sin^2 2\theta)$ parameter space under investigation and the fact that we model the flavor conversion physics more accurately.

\subsection{Feedback on the entropy and medium temperature}
\label{sec:entropy}

\begin{figure}[t]
\centering
\includegraphics[scale=0.3]{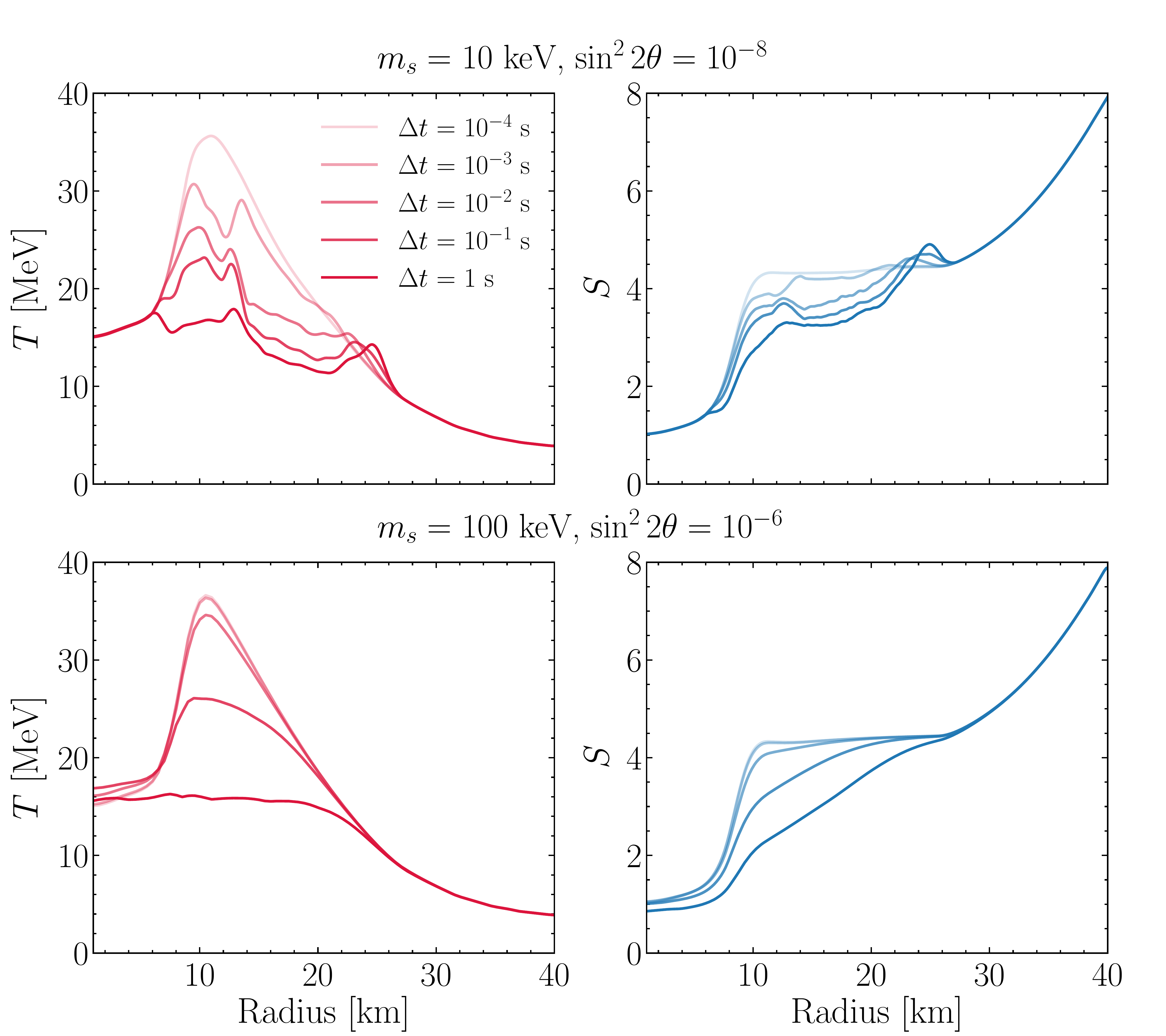}
\caption{Radial profile of the temperature (on the left) and the entropy per baryon (on the right) for  different times $\Delta t$, for  $(m_s, \sin^2 2\theta) = (10~\mathrm{keV}, 10^{-8})$  on the top panels and $(m_s, \sin^2 2\theta)  = (100~\mathrm{keV}, 10^{-6})$ on the bottom panels. The temporal evolution is represented through curves with different hues of the same color, from lighter to darker as $\Delta t$ increases. As a result of the dynamical feedback due to the production of sterile particles, $T$ and $S$ both decrease with time.} 
\label{fig:temperature}
\end{figure}

As outlined in Sec.~\ref{sec:equations}, the production of sterile particles and their reconversions  can channel energy from the interior of the proto-neutron star  to its outer parts, possibly affecting the proto-neutron star temperature.
Figure~\ref{fig:temperature} shows the radial profile of the medium temperature (on the left) and entropy per baryon (on the right) for  $(m_s, \sin^2 2\theta) = (10~\mathrm{keV}, 10^{-8})$  on the top panels and $(m_s, \sin^2 2\theta)  = (100~\mathrm{keV}, 10^{-6})$ on the bottom panels. Each quantity is shown for various $\Delta t$ to illustrate the temporal evolution. In agreement with the discussion in Sec.~\ref{sec:Beta}, the largest variations on the SN background occur between $10$ and $30$~km, i.e.~within the neutrino-sphere, and are negligible both deep inside in the proto-neutron star and in the proximity of $R_{\nu}$. On the top panels, as a result of the inefficient
collisional production of sterile particles in the SN core and because of the small neutrino degeneracy  after the temperature maximum (see Fig.~\ref{fig:Fig1}), more sterile antineutrinos are produced; as shown in Fig.~\ref{fig:feedback}, this leads to a negative change
of $Y_{\nu_s}$ as a function of time. The medium temperature and the entropy  both decrease as a function of time,
 because the neutrino cooling dominates in addition to the negative variation of $\Delta Y_{\nu_s}$ (see Fig.~\ref{fig:cooling} and Eq.~\ref{eq:Delta_entropy}).
Note that both the temperature and entropy increase around $\sim$~25~km due to the reconversion of sterile antineutrinos (see also Fig.~\ref{fig:feedback}).

The bottom panels of Fig.~\ref{fig:temperature} show a similar trend because sterile neutrinos are abundantly produced for $m_s=100$~keV; this induces a positive change in $Y_{\nu_s}$ that is responsible for a slower decrease of the temperature  when $\mu_{\nu_e}$ is still very large. Soon after, the neutrino cooling dominates and it determines the decrease of $T$ and $S$ as a function of time (see Fig.~\ref{fig:cooling} and Eq.~\ref{eq:Delta_entropy}). 

\begin{figure}[t]
\centering
\includegraphics[scale=0.35]{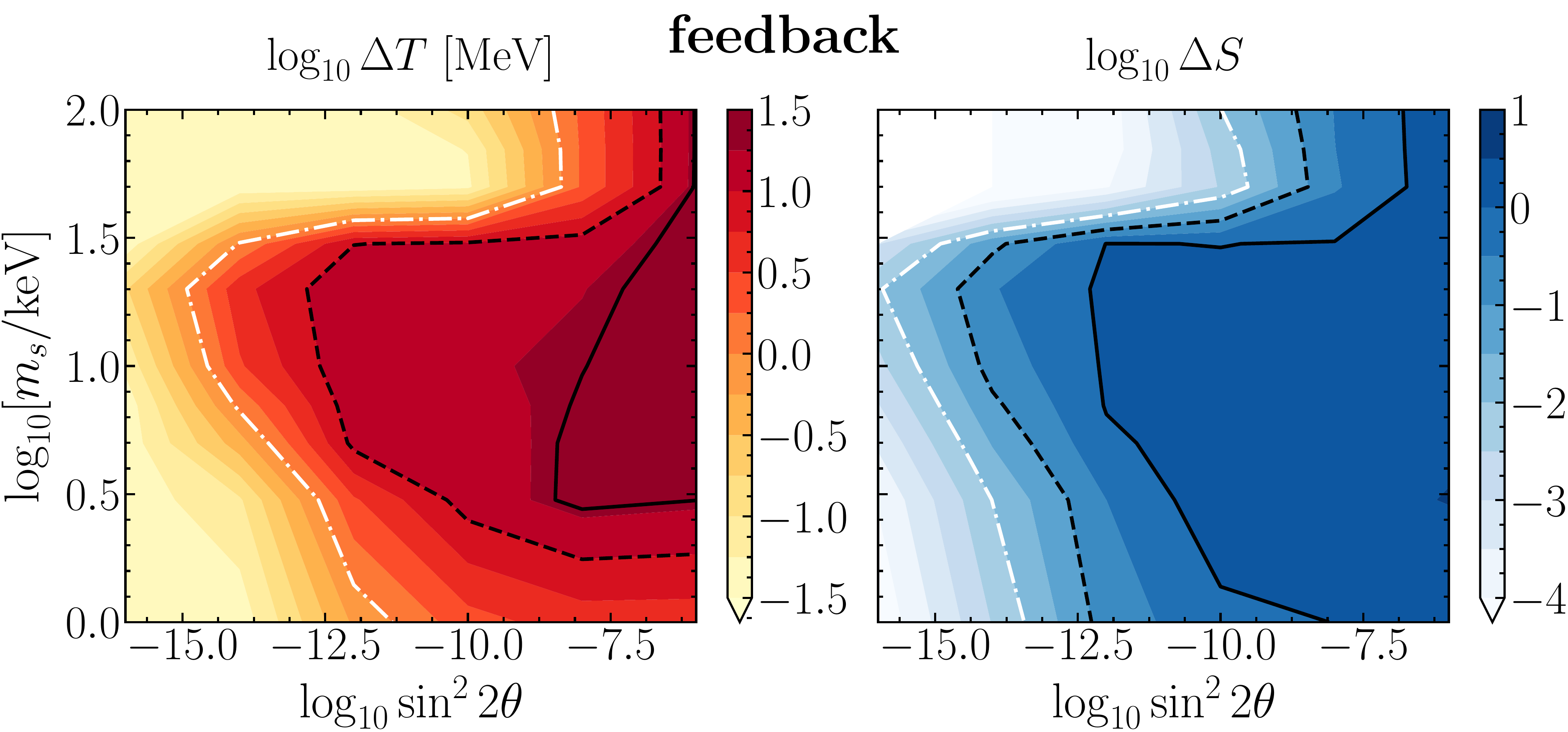}
\caption{Contour plot of the maximum difference between the initial ($\Delta t = 0~\mathrm{s}$)  and final ($\Delta t = 1~\mathrm{s}$, with dynamical feedback) medium temperature (on the left) and entropy per baryon (on the right) in the  ($\sin^2 2\theta, m_s$) parameter space. The black solid lines indicate $\Delta T > 20~
\mathrm{MeV}$ ($\Delta S > 1$), the black dashed lines represent $\Delta T > 10~\mathrm{MeV}$ ($\Delta S > 0.1$), and the white dash-dotted lines mark $\Delta T > 1~ \mathrm{MeV}$ ($\Delta S > 0.01$).  For large mixings, the variations in temperature and entropy are the largest due to the collisionally enhanced production of sterile particles and adiabaticity of the MSW resonant conversions. The maximum variations of temperature and entropy obtained for the scanned mixing neutrino parameter space are $\Delta T \simeq 25$~MeV and $\Delta S \simeq 3$.} 
\label{fig:entropy_contour}
\end{figure}
In order to generalize our findings to the  ($\sin^2 2\theta,  m_s$) parameter space, Fig.~\ref{fig:entropy_contour} illustrates the maximum variations in the medium temperature (left panel) and entropy per baryon (right panel), produced within the neutrinosphere. 
The  maximum variation in  temperature and entropy per baryon  is computed between the initial values of these quantities ($\Delta t=0$) and after  $\Delta t= 1$~s when the feedback effects are included.  The largest temperature variation is $\Delta T \simeq 25$~MeV and the largest variation of the entropy per baryon is $\Delta S \simeq 3$; both occur  for large mixing angles because, as the mixing angle increases, the conversions are more adiabatic and the number of sterile particles produced through collisions increases.

\newpage
\section{Implications on the supernova bounds on the mixing parameters}
\label{sec:SNbounds}
As stressed in Ref.~\cite{Suliga:2019bsq}, robust bounds on the sterile neutrino mass and mixing  from SNe can be obtained only through a self-consistent SN simulation where the microphysics related to the production of sterile particles is appropriately included. 
In order to place constraints on the allowed parameter space for  the sterile particles, we calculate the energy emitted in $\nu_s$ and $\bar\nu_s$ (see  Eq.~\ref{eq:Es}). This energy is then compared to the typical binding energy of a neutron star, $E_b = 3 \cdot 10^{53}~\mathrm{ergs}$. We caution the reader, that this comparison only provides a conservative estimate of the excluded mass-mixing parameter.

\begin{figure}[t]
\centering
\includegraphics[scale=0.35]{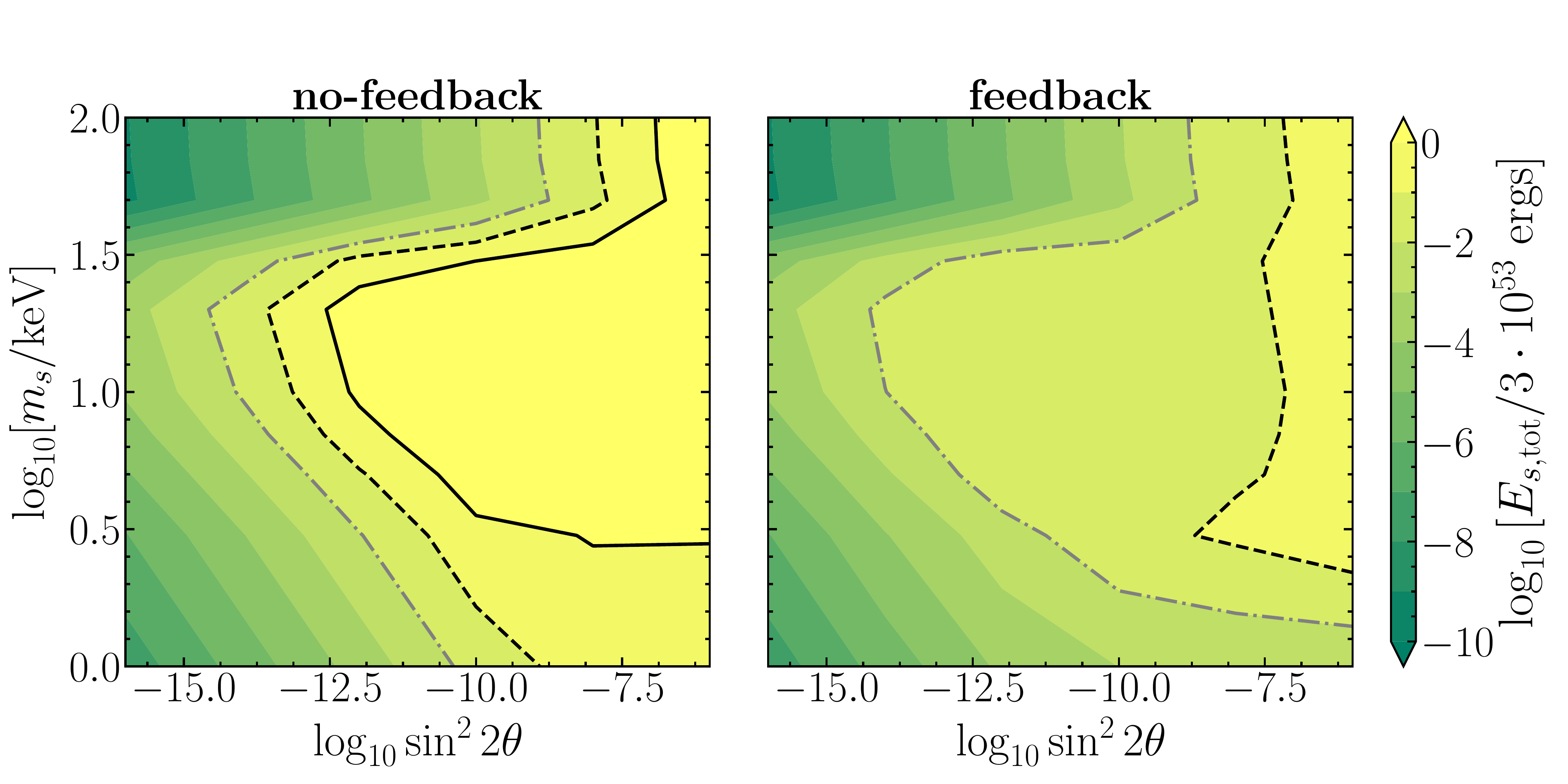}
\caption{Contour plot of the ratio of the total energy emitted in  sterile particles  without the dynamical feedback  due to the production of sterile particles  (left panel) and with dynamical feedback (right panel, $\Delta t = 1~\mathrm{s}$)  to the benchmark binding energy
($3 \cdot 10^{53}~\mathrm{ergs}$). The black lines indicate when the energy ratio is $> 1$ (solid) and $> 0.1$ (dashed); while the grey dash-dotted line indicates when the energy ratio is $> 0.01$.  No region of the parameter space is excluded from SNe when the dynamical feedback is included.} 
\label{fig:sterile_bounds}
\end{figure}
Figure~\ref{fig:sterile_bounds} shows the
exclusion contours on the mixing 
parameter space for the cases with (on the left) and without (on the right) dynamical feedback on the SN background due to the production of sterile particles. In agreement with the findings of Ref.~\cite{Suliga:2019bsq} for the $\tau$--$s$ mixing, the inclusion of the dynamical feedback considerably relaxes the bounds on the allowed sterile mass and mixing. Importantly, for the $e$--$s$ mixing, we do not find any excluded region ($E_{s,{\rm{tot}}}/E_b \ge 1$). 

Our results highlight the importance of a self-consistent radial and temporal dependent modeling of the sterile neutrino production in SNe. Although, we cannot draw roburst conclusions without running a self-consistent hydrodynamical simulation which includes sterile neutrinos, our results point towards the possibility that any $(m_s, \sin^2 2\theta)$ in the parameter space investigated in this work is compatible with the observation of the SN 1987a. 

\begin{figure}[t]
\centering
\includegraphics[scale=0.35]{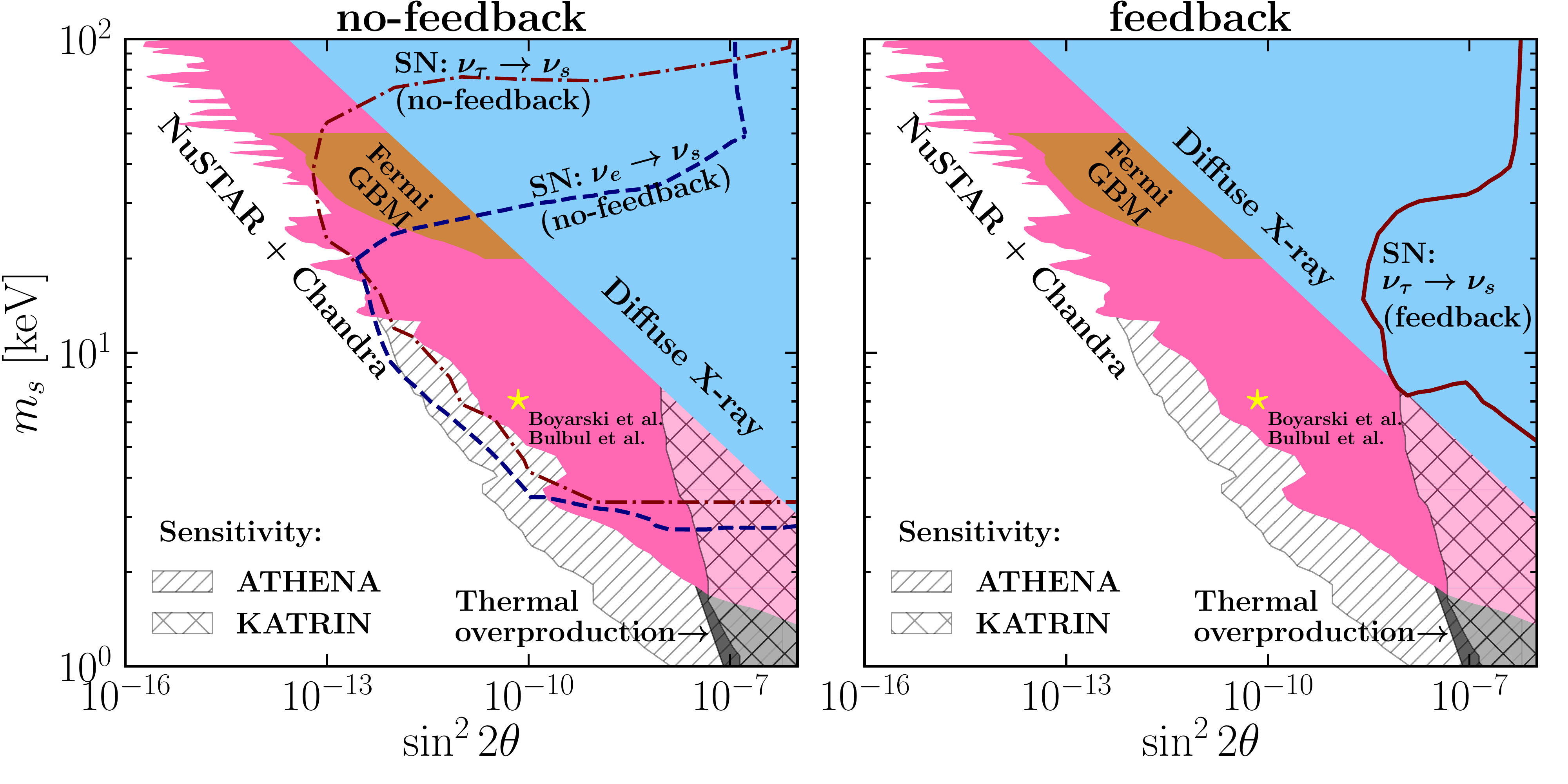}
\caption{Exclusion regions in the ($\sin^2 2\theta, m_s$) parameter space of sterile neutrino dark matter. Existing bounds include the X-ray constraints from observations of the M31 galaxy~\cite{Ng:2019gch,Horiuchi:2013noa} (NuSTAR$+$Chandra, in pink, for a more complete list of X-ray observational limits see, e.g.,~\cite{Boyarsky:2018tvu}), the diffuse X-ray background constraints~\cite{Boyarsky:2005us,Abazajian:2006jc} (in blue), the Fermi Gamma-Ray Burst Monitor all-sky spectral analysis~\cite{Ng:2015gfa} (in brown), the thermal overproduction through the Dodelson-Widrow mechanism~\cite{Dodelson:1993je,Abazajian:2017tcc} (in grey); the 3.5 keV line limit is plotted as a yellow star~\cite{Bulbul:2014sua,Boyarsky:2014jta}. The future sensitivities of KATRIN~\cite{Mertens:2018vuu} and ATHENA~\cite{Neronov:2015kca} are displayed as hatched regions.  For comparison, the SN exclusion region obtained without [with] dynamical feedback is plotted as a blue dashed (red dash-dotted) line for the $\nu_e-\nu_s$ ($\nu_\tau - \nu_s$~\cite{Suliga:2019bsq}) mixing on the left [right] panel. The ($\sin^2 2\theta, m_s$)  parameter space is unconstrained for the $\nu_e-\nu_s$ mixing (and almost unconstrained for the $\nu_\tau - \nu_s$ mixing) from SNe when the dynamical feedback due to the production of sterile particles is taken into account.}
\label{fig:sterile_dm}
\end{figure}

Our findings also affect the region of the parameter space considered in the context of dark matter searches, see e.g.~Refs.\cite{Merle:2017dhf,Boyarsky:2018tvu}. Figure~\ref{fig:sterile_dm} shows a summary of the  bounds on the allowed  ($\sin^2 2\theta, m_s$)  parameter space of sterile neutrino dark matter~\cite{Ng:2019gch,Ng:2015gfa,Horiuchi:2013noa,Boyarsky:2005us,Abazajian:2006jc,Neronov:2015kca},  the Dodelson-Windrow limit~\cite{Dodelson:1993je,Abazajian:2017tcc}, and the future sensitivity of  KATRIN~\cite{Mertens:2018vuu} and ATHENA~\cite{Neronov:2015kca}. For comparison, the exclusion regions computed for the $\nu_e-\nu_s$ mixing in this work and for the $\nu_\tau - \nu_s$ mixing in~\cite{Suliga:2019bsq} are plotted on the left (right) panel for the case without (with) dynamical feedback due to the production and propagation of sterile particles in the SN core.  Our results hint that SNe cannot exclude any region of the ($\sin^2 2\theta, m_s$) parameter space.

\section{Implications on  supernova explosions aided by sterile neutrinos}
\label{sec:SNexplosion}

According to the delayed neutrino-driven SN mechanism, neutrinos re-energize the stalled shock wave to
trigger the explosion~\cite{Burrows:2012ew,Janka:2006fh}. Sterile neutrinos with keV mass mixed 
with the electron flavors have been proposed as a possible ingredient to
aid the explosion~\cite{Hidaka:2006sg,Hidaka:2007se,Warren:2014qza,Warren:2016slz}. 
In fact, as discussed in Secs.~\ref{sec:MSW} and \ref{sec:Feedback}, the very-energetic sterile particles produced deep in the SN core may be reconverted
to active ones in the proximity of $R_\nu$ and therefore contribute to increase the amount of energy deposited by neutrinos in the gain layer. 

In order to derive a robust appraisal on whether sterile neutrinos can aid the SN explosion, a hydrodynamical SN simulation should be run self-consistently by including the sterile neutrino mixing, since the feedback on the SN characteristic quantities discussed in Sec.~\ref{sec:Feedback} should be explored within a dynamically evolving hydrodynamical background.
In this Section, we gauge the possibility that sterile neutrinos may aid the SN explosion by computing the amount of energy carried by  $\nu_e$'s and $\bar\nu_e$'s reconverted from sterile states in the outer part of the proto-neutron star  up to the neutrinosphere.

To this purpose, we compare the net deposited energy due to sterile neutrino production and reconversions in the outer part of the proto-neutron star between the $V_\mathrm{eff}$ minimum at $\Delta t=0$ (i.e., after $R~\sim~12~\mathrm{km}$) and $R_\nu$, to the gravitational  energy of this layer ($E_{G, \mathrm{out}}$), by defining
\begin{equation}
\label{eq:extra_energy}
\mathcal{R} = \frac{E_{G,\mathrm{out}} + E_{\nu_s \rightarrow {\nu_i}} - E_{\nu_s}}{E_{G, \mathrm{out}}} \ ,
\end{equation}
where $E_{\nu_s,\bar\nu_s}$ is the energy emitted in sterile neutrinos and antineutrinos, and $E_{\nu_s \rightarrow \nu_e, \bar\nu_s \rightarrow \bar\nu_e}$ is the energy reconverted from the sterile to the active sector from $R~\sim~12~\mathrm{km}$  to $R_\nu$.

\begin{figure}[t]
\centering
\includegraphics[scale=0.35]{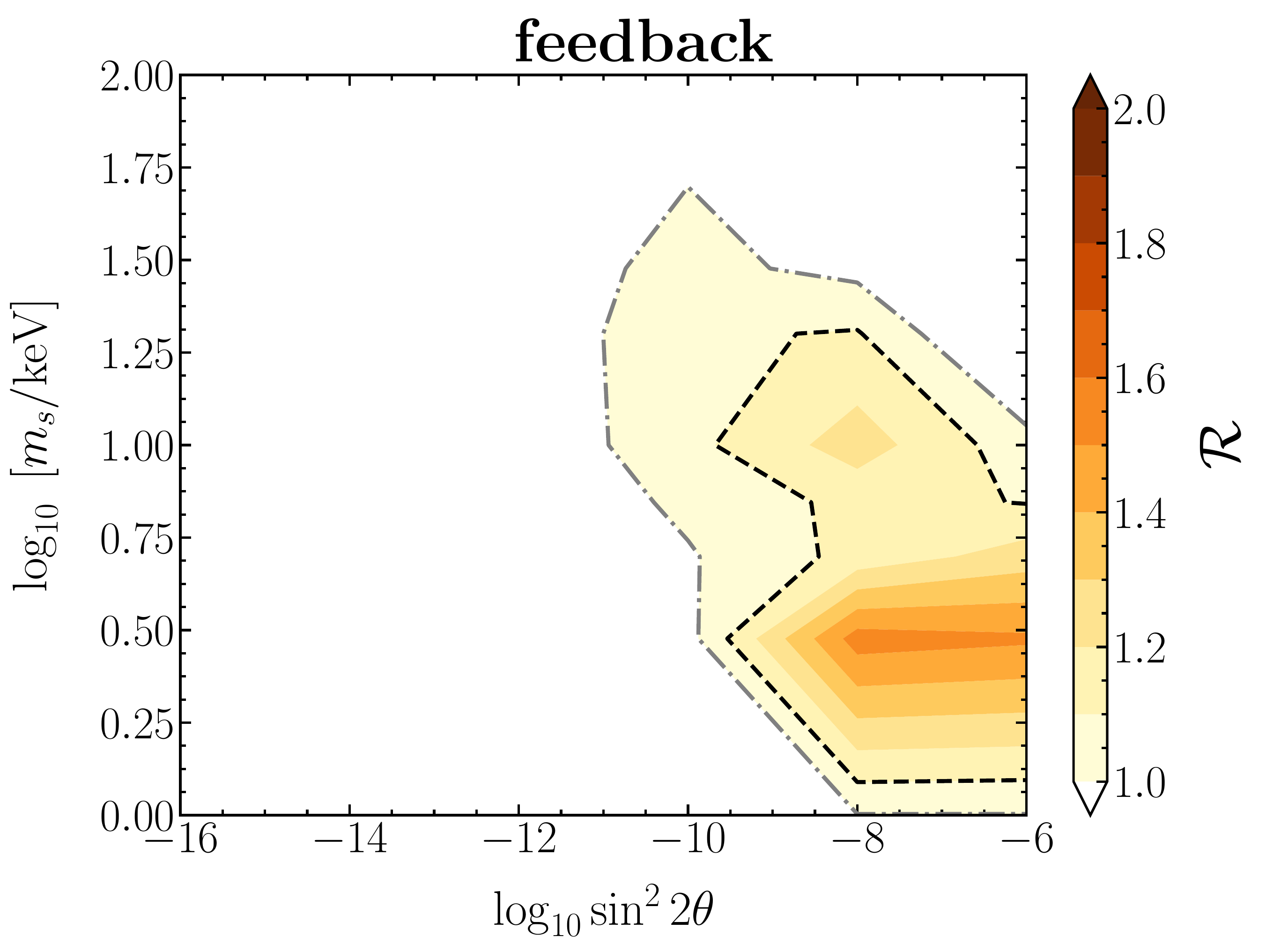}
\caption{Contour plot of the ratio $\mathcal{R}$ (see Eq.~\ref{eq:extra_energy}) 
 in the $(\sin^2 2\theta, m_s)$ parameter space. The black line indicates when the energy ratio is $\ge 1.1$ (dashed) and the grey dash-dotted line marks when the ratio is $\ge 1$. Only for a  small region of the parameter space with large $\sin^2 2\theta$, the reconversion of sterile particles into active flavors may aid the explosion. }
\label{fig:SNexplosion}
\end{figure}
Figure~\ref{fig:SNexplosion} shows the contour plot of the ratio $\mathcal{R}$ introduced in Eq.~\ref{eq:extra_energy} in the $(\sin^2 2\theta, m_s)$ parameter space. 
The net energy deposition in the outer layer of the proto-neutron star is positive ($\mathcal{R} \gtrsim 1$) only in a small region of the $(\sin^2 2\theta,  m_s)$
parameter space for large $\sin^2 2\theta$.
Outside this range, the net energy deposition is  negative ($\mathcal{R} \lesssim 1$).
For $m_s\simeq $ a few keV and $\sin^2 2\theta\gtrsim 10^{-8}$, the net energy deposition can reach up to $\sim 50\%$ of $E_{G,\rm out}$. This may potentially cause the outer part of the proto-neutron star to expand and enhance the emission of active neutrino flavor, as argued in Ref.~\cite{Warren:2014qza}.
We also note that, in our modeling, nearly all the reconverted $\nu_e$ and $\bar\nu_e$ are reabsorbed inside the neutrinosphere. Thus, the reconversion of sterile particles into active ones during the accretion phase do not directly contribute to the  heating.

\section{Conclusions}
\label{sec:Conclusions}

The mixing between electron and sterile neutrinos with mass between $1$ and $100$~keV is deemed to fundamentally affect the supernova mechanism. In this work, we shed new light on this issue by implementing the first complete, radial and time-dependent modeling of the mixing between electron and sterile neutrinos in the supernova core.

For simplicity, we rely on a static but radially evolving hydrodynamical background, typical of the supernova accretion phase, and explore the role of the dynamical feedback due to the production of sterile particles in the  $(m_s,\sin^2 2\theta)$ parameter space. We track the production of sterile particles by taking into account collisions in the supernova core as well as MSW resonant conversions.  Because of the shape of the effective matter potential, electron antineutrinos  undergo multiple MSW resonances in the supernova core. In addition, sterile neutrinos and antineutrinos may also be abundantly  produced through collisions. Reconversions from sterile into active flavors occur for the electron-sterile mixing and allow to reconvert very energetic sterile neutrinos produced in the innermost layers of the proto-neutron star  into active ones in the outermost layers of the proto-neutron star,   transporting energy from the interior of the proto-neutron star to its outer layers. 

In Ref.~\cite{Suliga:2019bsq}, we showed that the mixing of sterile with 
tau neutrinos may lead to the growth of a tau neutrino lepton asymmetry; where the latter is considerably overestimated when the dynamical feedback due to the production of sterile particles is not taken into account. In this work, we find that the electron-sterile  mixing  can modify not only the electron fraction but also
the chemical potentials of neutrons and protons. In addition, the electron-sterile mixing induces large variations on the entropy per baryon and on the supernova medium temperature.

For $\sin^2 2\theta \gtrsim 10^{-10}$, electron-sterile conversions may lead to dramatic  changes in the above mentioned supernova key quantities: the maximum electron fraction variation can be of $\simeq 0.21$ in the supernova core,  the medium temperature variation can reach $\simeq 25$~MeV, and a variation up to $3$ may occur for the entropy per baryon. 
These variations, due to the dynamical feedback, lead to a smaller amount of energy carried away by sterile neutrinos than previously estimated without taking into account the dynamical feedback.
Consequently, our findings suggest that a self-consistent appraisal of the impact of sterile neutrinos would lift the bounds on the electron-sterile mixing reported in the literature, leaving the parameter space relevant to dark matter searches unconstrained by the SN1987a cooling argument. 
However, we stress that a definitive answer can only be derived through a hydrodynamical simulation that consistently takes into account the sterile neutrino transport and the flavor conversion physics.

We also explored whether the energy transport, due to the reconversion of sterile (anti)neutrinos to electron (anti)neutrinos, from the interior of the proto-neutron star to its outer layers  can aid the supernova explosion. For most of the parameter space considered in this work, this effect is negligible. The only exception is a small region of the parameter space ($m_s\sim$~few keV and $\sin^2 2\theta\gtrsim 10^{-8}$) where the energy transport  may be considerably altered by the sterile neutrino mixing.

In conclusions, sterile neutrinos with mass $1$--$100$~keV may have a major impact on the supernova physics. Our work highlights the relevance of the dynamical feedback due to the production of sterile particles on the supernova mechanism and the importance of consistently modeling the sterile flavor conversions within a radial and time-dependent supernova model. Only through a self-consistent assessment of the sterile neutrino conversion physics, one could finally assess the role of these particles in the supernova mechanism.

\acknowledgments
We are grateful to Robert Bollig and Thomas Janka for granting access to the data of the supernova model adopted in this work.
This work was supported by the Villum Foundation (Project No.~13164),  the Carlsberg Foundation (CF18-0183), the Knud H\o jgaard Foundation, the Deutsche Forschungsgemeinschaft through Sonderforschungbereich SFB 1258 ``Neutrinos and Dark Matter in Astro- and Particle Physics'' (NDM), the Academia Sinica under Grant No. AS-CDA-109-M11, the Ministry of Science and Technology, Taiwan under Grant No. 108-2112-M-001-010, and the Physics Division, National Center of Theoretical Science of Taiwan.

\appendix
\section{Charged current interaction rates}
\label{appendix:Pauli}
Electron neutrinos and antineutrinos undergo CC and NC interactions in the dense SN core. We introduced the NC interaction rates in Appendix A of Ref.~\cite{Suliga:2019bsq}, while focus on the CC interactions in this appendix. 

The main CC processes affecting the electron flavors are the following: 
\begin{equation}
\label{eq:beta_reac1}
\mathrm{e^-} + \mathrm{p} \leftrightarrow \nu_e + \mathrm{n} \quad \mathrm{and} \quad \mathrm{e^+} + \mathrm{n} \leftrightarrow \bar{\nu}_e + \mathrm{p} \ .
\end{equation}
The correspondent interaction rates are (see also Eq.~\ref{eq:coll_rate})

\begin{equation}
\Gamma(\nu_e + \mathrm{n}  \rightarrow \mathrm{e^+} + \mathrm{p}) = n_B Y_\mathrm{n} \frac{\int dE F_{\rm{CC}, n}(E) \sigma_{\nu \mathrm{n, CC}}(E) dn_{\nu_e}/dE}{n_{\nu_e}}\ ,
\end{equation}
\begin{equation}
\Gamma(\bar\nu_e + \mathrm{p}  \rightarrow \mathrm{e^-} + \mathrm{n}) = n_B Y_\mathrm{p} \frac{\int dE F_{\rm{CC}, p}(E) \sigma_{\nu \mathrm{p, CC}}(E) dn_{\bar\nu_e}/dE}{n_{\bar\nu_e}}\ ,
\end{equation}
where the CC cross section $\sigma_{\nu p,n}$ is defined as in Ref.~\cite{Strumia:2003zx}, $dn_{\nu}/dE$ is the neutrino energy distribution introduced in Sec.~\ref{sec:ref_signal}, $Y_{p (n)}$ is the proton (neutron) fraction,  $n_B$ is the baryon density, and $F_{\mathrm{CC}, n,p}(E)$ is the Pauli blocking factor estimated in the following.

\begin{figure}[t]  
\centering
\includegraphics[scale=0.35]{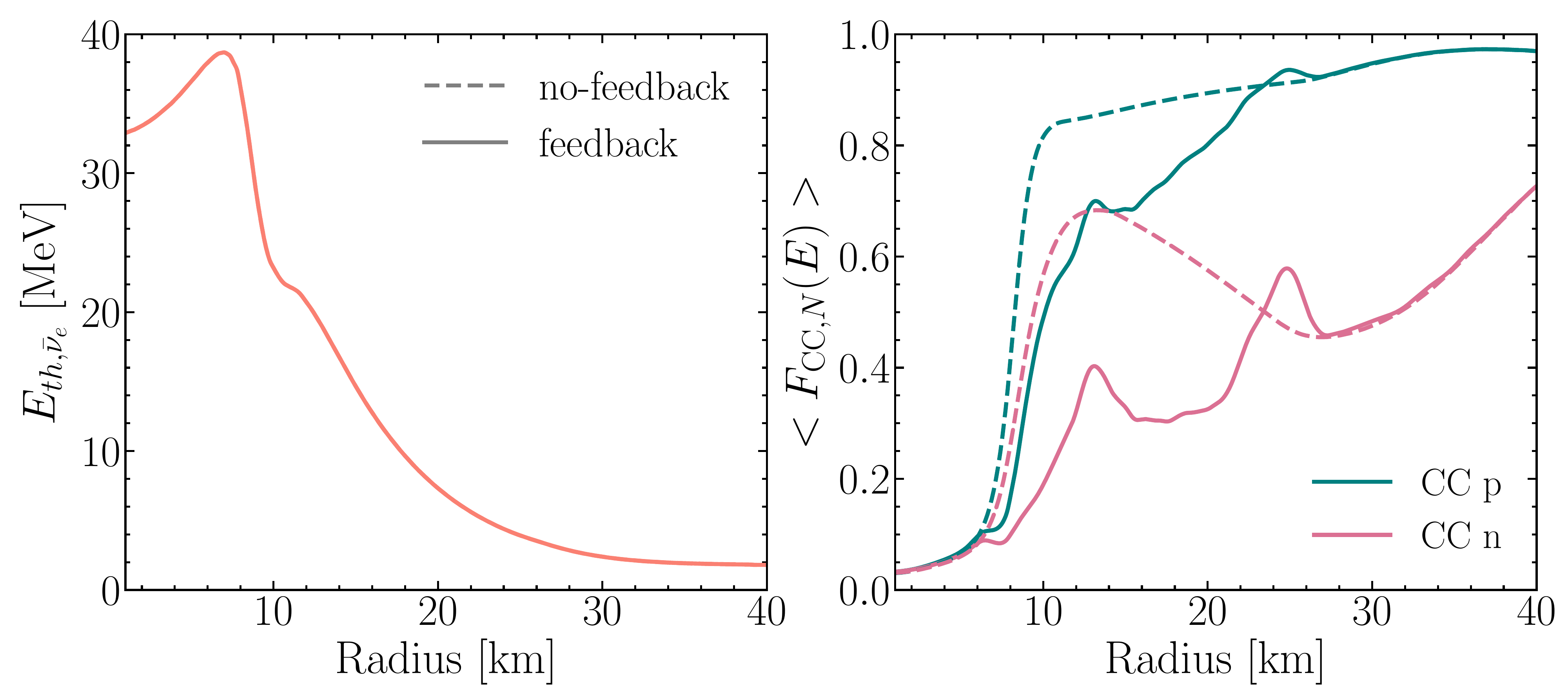}
    \caption{Radial profile of the antineutrino threshold energy, $E_{{\rm{th}}, \bar \nu_e}$ (left panel) and of the energy averaged Pauli blocking factors for the $\beta$ reactions (right panel). The solid (dashed) lines show the results with (without) the dynamical feedback effects for $(m_s, \sin^2 2\theta) = (10~\mathrm{keV}, 10^{-8})$. The change in the Pauli blocking factors follows the change in $Y_e$.}
    \label{fig:Pb}
\end{figure}

Given the high density present in the SN core, not all final states of a given reaction are allowed because of Pauli blocking~\cite{Raffelt:1996wa,Lamb:1976ac}. As a consequence of Pauli blocking, the total interaction rate is suppressed at high densities.
We calculate the Pauli blocking factors for the CC scattering following the method employed in Refs.~\cite{Raffelt:1996wa,Bruenn:1985en,Suliga:2019bsq}. The Pauli blocking factor for the electron antineutrino CC reaction is
\begin{equation}
\label{eq:Pauli_blocking_CC_nu_e_bar}
F_\mathrm{CC, p} (E) = \left( 1 - f(E_{e^{+}}) \right) \int_{0}^{\infty} \frac{2}{2 \pi^2 n_{\mathrm{p}}} {dp p^2} f_{\mathrm{p}}(E_{\mathrm{p}}) \left( 1 - f_{\mathrm{n}}(E_\mathrm{n}) 	\right) \ ,
\end{equation} 
where $E_{\mathrm{p, n}} = \sqrt{p^2 + m^2_{\mathrm{p,n},\star}} + U_{\mathrm{p,n},\star}$ is the nucleon energy, $p$  the  momentum of the nucleon,  $m_{\mathrm{p,n},\star}$ the effective nucleon mass, $U_{\mathrm{p,n},\star}$ the nucleon mean field potential, and $f_{n,p}(E_{n,p})$ the Fermi-Dirac energy distribution of nucleons. The neutrino energy is expressed in terms of  the positron energy through the following relation
\begin{equation}
\label{eq:positron_momentum1}
p_{e^+} = \sqrt{\left( \frac{2E (m_\mathrm{p} + U_\mathrm{p})+  2{U_\mathrm{p}}m_\mathrm{p} +U_\mathrm{p}^2+ m_\mathrm{p}^2 - m_\mathrm{n}^2 - m_{e^+}^2  - U_{\mathrm{n}}^2  -2 U_\mathrm{n} m_\mathrm{n} }{2 (m_\mathrm{n} + U_\mathrm{n} )}  \right)^2 - m_{e^+}^2}\ ,
\end{equation}
by taking into account that  $E_{e^+}= \sqrt{p_{e^+}^2 + m_{e^+}^2}$.
The $\bar\nu_e$ threshold energy derived from Eq.~\ref{eq:positron_momentum1} is
\begin{equation}
E_{\mathrm{th}} =  
\frac{2m_{e^+}(U_\mathrm{n} + m_\mathrm{n}) + U_\mathrm{n}^2 + 2 U_\mathrm{n} m_\mathrm{n} + m_\mathrm{n} + m_{e^+} - m_\mathrm{p}^2 - U_\mathrm{p}^2 - 2U_\mathrm{p} m_\mathrm{p}}{2(m_\mathrm{p} + U_\mathrm{p})}\ .
\end{equation}
Note that the threshold energy deviates from the standard value, $E_{\mathrm{th}} \approx 1.8~\mathrm{MeV}$ for non-degenerate nucleons, because of the nucleon mean-field potentials arising in extreme conditions of the SN core. The radial profile of the $\bar\nu_e$ threshold energy is shown in the left panel of Fig.~\ref{fig:Pb}.

Analogously, for the  CC scattering of electron neutrinos, the Pauli blocking factor is given by 
\begin{equation}
F_{\rm{CC},n} (E) = \big(1-f(E_{e^-})\big) \int_{0}^{\infty} \frac{2}{2 \pi^2 n_{\mathrm{n}}} {dp p^2} f_{\mathrm{n}}(E_{\mathrm{n}}) \big(1-f_{\mathrm{p}}(E_\mathrm{p})\big) \ ,
\end{equation} 
where $E_{e^-}= \sqrt{p_{e^-}^2 + m_{e^-}^2}$ is the electron energy and the electron momentum is
\begin{equation}
\label{eq:electron_momentum}
p_{e^-} = \sqrt{\left(\frac{U_\mathrm{n}^2 + 2E(m_\mathrm{n} + U_\mathrm{n}) + m_\mathrm{n}^2  + 2 U_\mathrm{n} m_\mathrm{n} - U_\mathrm{p}^2 - 2 U_\mathrm{p} m_\mathrm{p} - m_\mathrm{p}^2 - m_{e^-}^2}{2(m_\mathrm{p} + U_\mathrm{p} )}\right)^2 - m_{e^-}^2} \ .
\end{equation}

The right panel of Fig.~\ref{fig:Pb} shows the radial profile of the energy averaged Pauli blocking factors for the reactions from Eq.~\ref{eq:beta_reac1}. The dashed lines represent the Pauli blocking factors for calculations without the dynamical feedback, whereas the solid lines reflect how the factors change after $\Delta t =1$~s  including the dynamical feedback, as described in the Sec.~\ref{sec:Feedback}, for  $(m_s, \sin^2 2\theta) = (10~\mathrm{keV}, 10^{-8})$. In the region affected by the sterile neutrino production $\langle F_{\mathrm{CC, p(n)}}(E) \rangle$ decreases (increases) following the $Y_e$ increase (decrease). 
Note that, in our work, we rely on the Pauli blocking factors averaged over the neutrino energy distribution. The difference between the energy averaged Pauli blocking and its  energy-dependent expression has a negligible impact on the final results, but it allows to gain in computational time.

\begin{figure}[t]  
\centering
\includegraphics[scale=0.4]{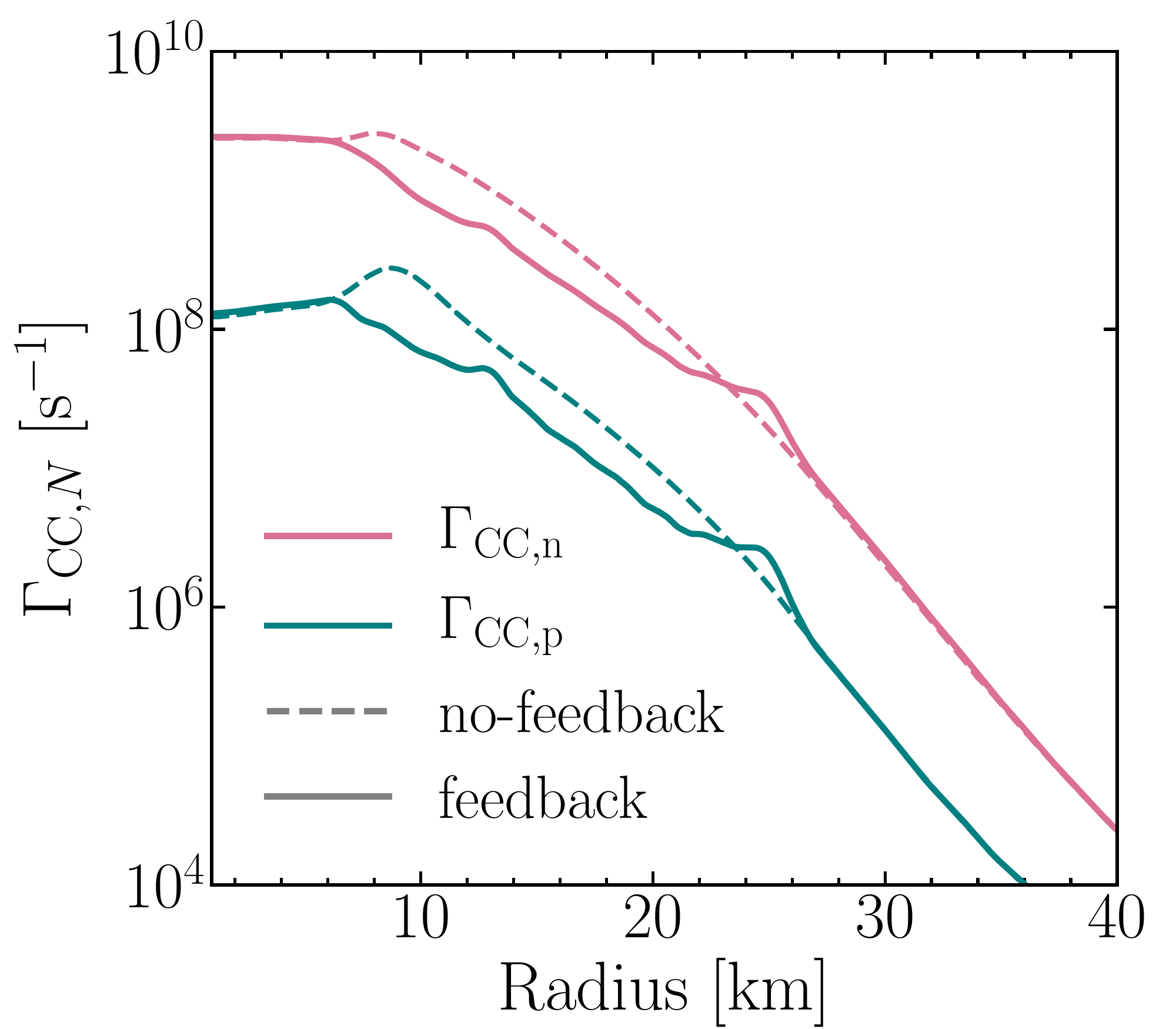}
    \caption{Radial profile of the charged-current interaction rates, $\Gamma_{\nu_e}$ (teal) and $\Gamma_{\bar\nu_e}$ (pink), averaged over the  (anti-)neutrino energy distribution. The feedback case (solid lines) is calculated for $(m_s, \sin^2 2\theta) = (10~\mathrm{keV}, 10^{-8})$ after $\Delta t = 1$~s. The rates decrease (increase) because of the decrease (increase) in the medium temperature caused by the sterile neutrino and antineutrino production.}
    \label{fig:rates}
\end{figure}

Figure~\ref{fig:rates} shows the radial profile of the CC interaction rates obtained by averaging over the neutrino energy spectrum for  $\nu_e$  and $\bar\nu_e$ and taking into account the Pauli blocking factors. One can see that the dynamical feedback (see Sec.~\ref{sec:Feedback}) is responsible for inducing a variation of the interaction rates due to the dynamical feedback on the medium temperature caused by the production of sterile particles.

In our work, we take into account the replenishment of electron neutrinos and antineutrinos through the electron and positron capture reactions. When  the production rate of the sterile (anti-)neutrinos is higher than the (positron) electron capture rate, the former limits the production.

\section{Heating and cooling rates due to the production of sterile particles}
\label{sec:app_qcdot}

For the heating rate used in Sec.~\ref{sec:Heating} due to the reconversion of sterile (anti)neutrinos, we follow Ref.~\cite{Nunokawa:1997ct} and define it as
\begin{equation}
\label{eq:heating_rate}
Q^{h}_{\nu_s}(r, t) = \sum_{\nu = \nu_e, \bar{\nu}_e} \frac{\dot{E}^{h}_{\nu} (r, t)}{N_B(r)} \,
\end{equation}
where the number of baryons in a single SN shell is $N_B(r_i) = V(r_i) \: n_B(r_i) $.
The energy gain coming from the sterile neutrino or antineutrino ($\nu = \{ \nu_e, \bar{\nu}_e \}$) reconversions, per unit time is
\begin{gather}
\label{eq:energy_gain_sterile}
\begin{aligned}
\dot{E}^{h}_{\nu} (r, t) \sim {}&\sum_{k=1}^L \left[ P_{\mathrm{se}}(E_k, r, t) \Theta \left(\frac{\Delta r}{\lambda_\nu (E_k, r)}\right) \sum_{j=1}^{i-1}  P_{\mathrm{es}}(E_k, r_j, t) \: \frac{dn_\nu}{dE}(r_j, t) \:  \frac{r_j^2}{r_i^2} \ dE_k  E_k \right] \\ & \times V(r) \Delta r^{-1} \ ,
\end{aligned}
\end{gather}
where the function $\Theta$  determines the amount of energy deposited in the SN shell
\begin{equation}
\label{eq:Heaviside}
\Theta (x) = \begin{cases} 1 &\mbox{if } x \geq 1 \\
x & \mbox{if } x < 1 \end{cases} \ . 
\end{equation}
If the width of the SN shell is smaller than the mean-free path of the active neutrino $\Delta r / \lambda_\nu < 1$, all energy reconverted to the active neutrinos cannot be transferred back to the medium within the SN shell of width $\Delta r$. 
In this case, only a fraction of the reconverted energy is deposited in the medium in the SN shell. The remaining part is propagated to the next neighbouring shells and redistributed among them accordingly.

The energy loss per baryon due to the sterile neutrino production is computed similarly to the heating rate, 
\begin{equation}
\label{eq:cooling_rate}
Q^{c}_{\nu_s}(r, t) = \sum_{\nu = \nu_e, \bar{\nu}_e} \frac{\dot{E}^{c}_{\nu} (r, t)}{N_B(r)} \ ,
\end{equation}
where the energy loss per baryon due to the sterile neutrino production
\begin{gather}
\label{eq:energy_gain_cooling}
\begin{aligned}
\dot{E}^{c}_{\nu} (r, t) \sim  V(r) {\Delta r}^{-1} \sum_{k=1}^L P_{\rm{es}}(E_k, r, t) \ \frac{dn_{\nu}}{dE_k} (r, t) \ dE_k E_k  \ .
\end{aligned}
\end{gather}

\begin{figure}[t]
\centering
\includegraphics[scale=0.4]{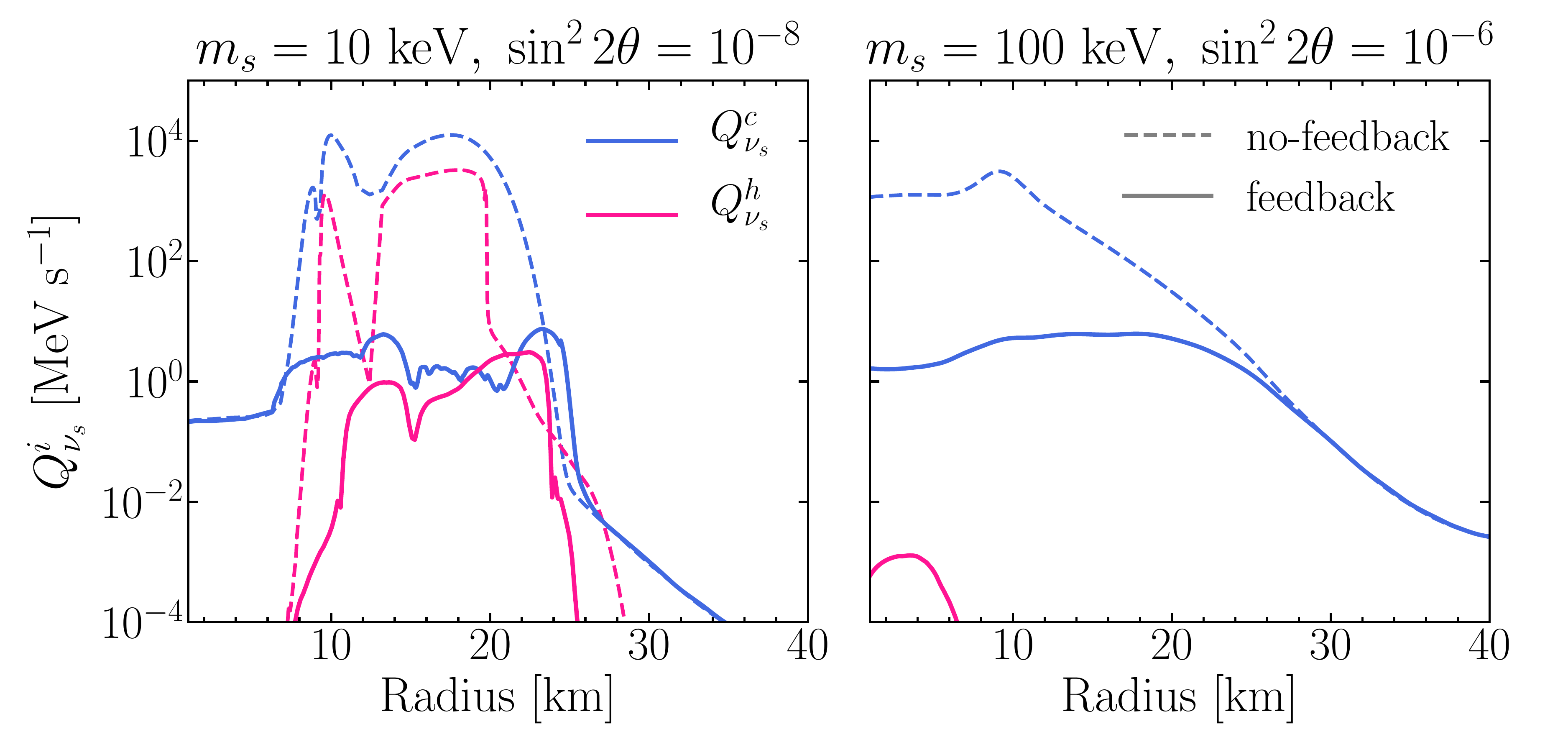}
\caption{Radial profile of the neutrino heating (magenta) and cooling rates (blue) defined as in Eqs.~\ref{eq:heating_rate} and \ref{eq:cooling_rate} for $(m_s, \sin^2 2\theta)= (10~\mathrm{keV}, 10^{-8})$ on the left panels and $(m_s, \sin^2 2\theta)= (100~\mathrm{keV}, 10^{-6})$ on the right. The dashed (continuous) lines represent the cases without (with) dynamical feedback. As $\sin^2 2\theta$  increases, the amount of cooling increases. The heating for the $(m_s, \sin^2 2\theta)= (10~\mathrm{keV}, 10^{-8})$ case is comparable to the cooling, while it is negligible for the $(m_s, \sin^2 2\theta)= (100~\mathrm{keV}, 10^{-6})$ case because no MSW resonances occur for $\Delta t=0$ and only very few high-energy energy modes undergo MSW conversions as $\Delta t$ increases.} 
\label{fig:cooling}
\end{figure}

In order to compare $Q^{h}_{\nu_s}$ and $Q^{c}_{\nu_s}$, Fig.~\ref{fig:cooling} shows their radial profile for $(m_s, \sin^2 2\theta)= (10~\mathrm{keV}, 10^{-8})$ on the left and $(m_s, \sin^2 2\theta)= (100~\mathrm{keV}, 10^{-6})$ on the right. 
 For both mixings, the neutrino cooling dominates over the heating for most radii. For the $10~\mathrm{keV}$ case, the main source of cooling comes from the MSW conversions of  antineutrinos, with the collisional production taking over at very small  ($r \lesssim 6~\mathrm{km}$) and large distances from the core  ($r \gtrsim 26~\mathrm{km}$). This happens because very few energy modes can undergo MSW resonances in these regions. Nevertheless, there is a small radial region, initially around $r\sim 25~\mathrm{km}$, where the heating stemming from the $\bar\nu_s$ reconversions dominates over the the cooling. (This also causes entropy and temperature to rise, as can be seen in Fig.~\ref{fig:temperature}.) With time, the sterile neutrino cooling and heating decline, as a result of the less efficient production of $\nu_s$ and $\bar\nu_s$ due to the feedback effect.
The dip  at $r \simeq 12~\mathrm{km}$, which originates from the effective potential minimum at that point, disappears at $\Delta t$ increase because of the $V_\mathrm{eff}$ evolution. For the $m_s =100~\mathrm{keV}$ case, the cooling exceeds the heating for all radii and times. However, contrary to the $m_s =10$~keV case,  the collisional production of neutrinos is the main responsible for the cooling. As $\Delta t$ increases, the heating gradually increases, as the effective potential decreases, and allows the energy modes below $1000$~MeV to undergo resonantly enhanced conversions. In spite of that, due to the decreasing temperature, the high electron neutrino chemical potential, and the location of the MSW region in the proximity of the SN core, the amount of the produced and the reconverted $\bar\nu_s$ is negligible.

\section{Interpolation procedure for the medium temperature}
\label{appendix:Temperature}

Section~\ref{sec:entropy} focuses on the computation of the effect of the dynamical feedback due to the production of sterile particles on the entropy and medium temperature. For fixed baryon density, the feedback effects have been implemented through the CompOSE package~\cite{Typel:2013rza}.

At each time step and radius, the medium  temperature is updated through the CompOSE tables  by varying the entropy per baryon, and fraction of electrons.
\begin{figure}[t]   
\centering
\includegraphics[scale=0.35]{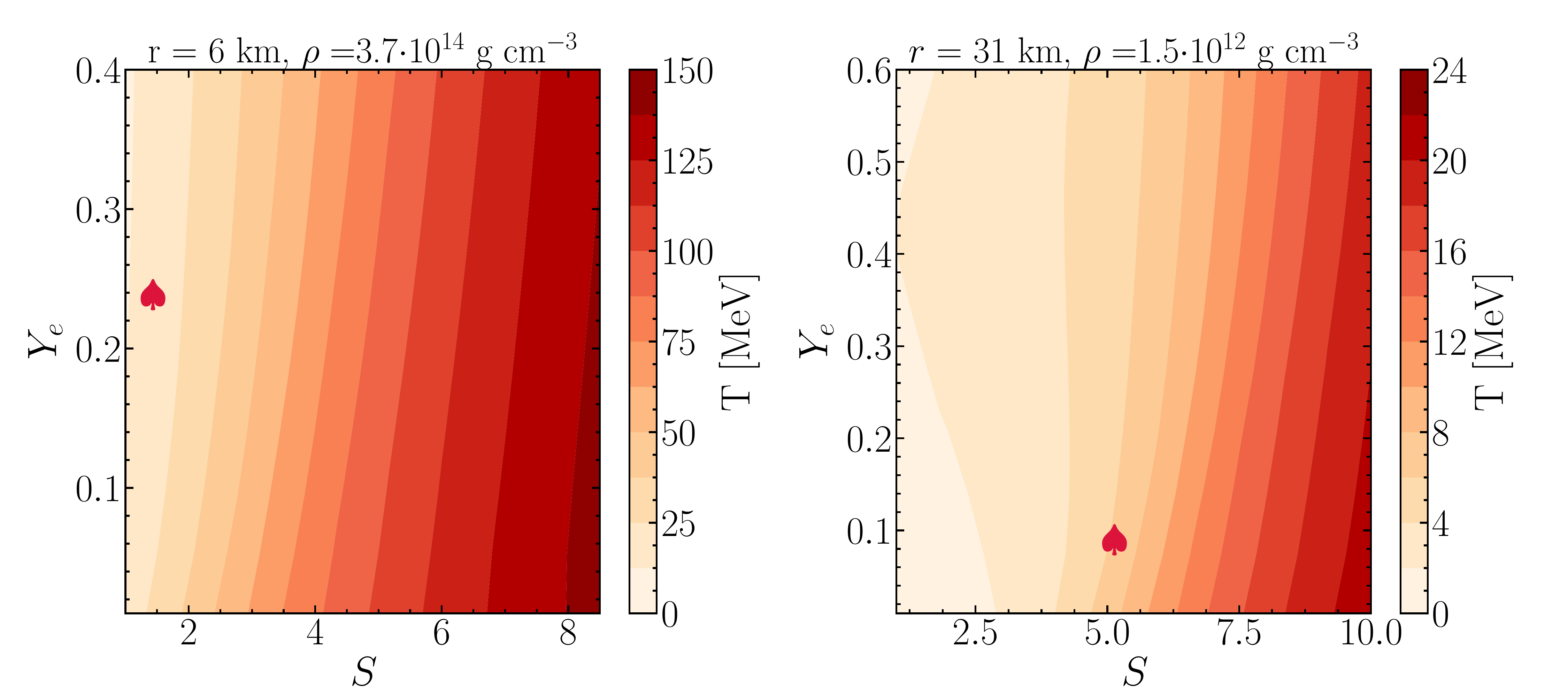}
    \caption{Contour plot of the medium temperature in the plane spanned by the entropy and the electron fraction a baryon density of $3.7 \cdot 10^{14}$~g cm$^{-3}$ on the left and for $1.5 \cdot 10^{12}$~g cm$^{-3}$ on the right. The SN background adopted for both panels is the one at $t_\mathrm{pb} = 0.25~\mathrm{s}$. The $\spadesuit$ symbols indicate the initial values of $Y_e$ and $S$ as  from the hydrodynamical simulation.}
    \label{fig:Temperature_interpolation}
\end{figure}
Figure \ref{fig:Temperature_interpolation} provides an example of the interpolation procedure for two different values of the baryon density $\rho$ in the plane spanned by   $Y_e$ and $S$. The $\spadesuit$ symbols indicate the initial configuration of the SN simulation for $t_\mathrm{pb} = 0.25~\mathrm{s}$. For fixed $\rho$, by computing $Y_e$ and $S$ through the procedure illustrated in Sec.~\ref{sec:entropy} to include the feedback due to the production of sterile particles, one obtains a new pair of $(S,Y_e)$ which correspond to a new value for the medium temperature $T$. This procedure has been adopted to estimate the medium temperature as a function of the radius and in time.

\bibliographystyle{JHEP}
\bibliography{sterile_electron}
\end{document}